\documentstyle[12pt,psfig]{article} 
\begin{document} 
\newcommand{\cd}{\makebox[0.08cm]{$\cdot$}} 
\centerline{\bf Triple-Quark Elastic Scatterings and Thermalization} 
\vskip 18pt 
\centerline{Xiao-Ming X${\rm u}^{{\rm a},{\rm b}}$,  
Peng R${\rm u}^{\rm c}$, H. J. Webe${\rm r}^{\rm d}$} 
\vskip 14pt 
\centerline{$^{\rm a}$Department of Physics, Shanghai University, Baoshan,  
Shanghai 200444, China} 
\centerline{$^{\rm b}$Nuclear Physics Division, Shanghai Institute of Applied  
Physics} 
\centerline{Chinese Academy of Sciences, P.O.Box 800204, Shanghai 201800, 
            China} 
\centerline{$^{\rm c}$Department of Physics, Wuhan University of Science and 
Technology, Wuhan 430081, China} 
\centerline{$^{\rm d}$Department of Physics, University of Virginia,  
Charlottesville, VA 22904, U.S.A.}

\begin{abstract} 
\baselineskip=14pt 
Triple-quark elastic scattering amplitudes from perturbative QCD are first
calculated and then used 
in a transport equation to study the thermalization of quark matter. 
By examining momentum isotropy to which the transport equation leads, we can
determine thermalization time and offer an initial thermal quark distribution
function.
With an anisotropic initial quark distribution, which is relevant to quark 
matter initially created in a central Au-Au collision at $\sqrt {s_{NN}}=200$ 
GeV, the transport equation gives a time of the order of 1.8 fm/$c$ for quark
matter itself to thermalize by the triple-quark scatterings.
\end{abstract} 
\leftline{PACS codes: 24.85.+p; 12.38.Mh; 25.75.Dw; 25.75.Gz} 
\leftline{Keywords: Triple-quark scattering; transport equation;  
thermalization} 

\newpage 
\leftline{\bf 1. Introduction} 
\vspace{0.5cm} 
Hydrodynamic calculations [1-7] that explain the elliptic flow coefficient  
${\rm v}_2$ [8,9] at $p_{\bot} < 2$ GeV/$c,$ lead to the conclusion that  
quark-gluon matter created in Au-Au collisions at the Relativistic Heavy Ion  
Collider (RHIC) thermalizes rapidly. The usual parton scatterings give a long 
thermalization time of quark-gluon matter [10]. However, triple-gluon  
elastic scatterings in gluon matter lead to a short thermalization time  
of the order of 0.45 fm/$c$ [11]. Such triple-gluon elastic scatterings can  
occur in view of the high gluon number density that has been achieved in Au-Au 
collisions [12-14]. However, since quark-gluon matter includes quarks and  
measured quantities rely on the presence of quarks, it is important to  
investigate the thermalization of quark matter based on initial anisotropic 
quark  
distributions. At present, the thermalization mechanism of quark matter is  
unknown. In this letter, we study the thermalization based on two- and  
three-quark elastic scattering processes. Until now, 
the squared amplitudes for the triple-quark scatterings have not been
studied and must be calculated in perturbative QCD. Therefore, this work 
comprises two new tasks: one is to derive the squared amplitudes and the other
is to study the thermalization.
In order to keep the physics  
relating to the triple-quark elastic scatterings transparent, and due to the  
complications of calculating the squared amplitudes and numerically solving a  
transport equation, we exclude here the quark scatterings with gluons and/or  
antiquarks.  

\vspace{0.5cm} 
\leftline{\bf 2. Quark-quark and triple-quark scatterings} 
\vspace{0.5cm} 
The evolution of quark matter relies on the scattering of two quarks with  
identical or different flavors. Expressed in terms of the Mandelstam variables
($s,t,u$), the spin- and color-averaged squared amplitudes of order  
$\alpha_{\rm s}^2$ in perturbative QCD [15,16] are  
\begin{equation} 
\mid {\cal M}_{{\rm uu} \to {\rm uu}} \mid^2 
=\mid {\cal M}_{{\rm dd} \to {\rm dd}} \mid^2 
={\rm g}_{\rm s}^4 \bigg [ \frac {4}{9} \bigg ( \frac {s^2+u^2}{t^2}  
+ \frac {s^2+t^2}{u^2} \bigg ) -\frac {8}{27} \frac {s^2}{ut} \bigg ] 
\end{equation} 
for the scattering of two up (down) quarks and 
\begin{equation} 
\mid {\cal M}_{{\rm ud} \to {\rm ud}} \mid^2 
={\rm g}_{\rm s}^4 \frac {4}{9} \frac {s^2+u^2}{t^2} 
\end{equation} 
for the scattering of up and down quarks. Here, ${\rm g}_{\rm s}$ is  
the quark-gluon coupling constant and ${\rm g}_{\rm s}^2 =4\pi \alpha_{\rm s}$.

The diagrams for elastic scattering processes of three quarks at order  
$\alpha_{\rm s}^4$ are drawn in Figs. 1 and 2. The three-quark scattering  
processes in Fig.~1 and Fig.~2 are separated into two distinct classes of  
diagrams:\\(1)~${\rm A}_-, {\rm A}_{-45}, {\rm A}_{-56}, {\rm A}_{-64},  
{\rm A}_{-4(56)}, {\rm A}_{-(45)6}$ of Fig.~1\\ 
involving two successive gluon exchanges and,\\ 
$(2)~{\rm A}_*, {\rm A}_{*45}, {\rm A}_{*56}, {\rm A}_{*64}, {\rm A}_{*4(56)}, 
{\rm A}_{*(45)6}$ in Fig.~2,\\involving a triple gluon coupling. 
  
When two final quarks have the same flavor, an exchange of the two quarks 
generates a different diagram from those diagrams $\rm {A}_-$ in Fig.~1 or  
$\rm {A}_*$ in Fig.~2. When three quarks have the same flavor, exchanges of the
 three  
quarks lead to the diagrams ${\rm A}_{-4(56)}$ and ${\rm A}_{-(45)6}$ of Fig.~1
and  ${\rm A}_{*4(56)}$ and ${\rm A}_{*(45)6}$ of Fig.~2. 

A quark four-momentum is labeled as $p_{\rm i}=(E_{\rm i},\vec {p}_{\rm i})$  
in the process ${\rm q}(p_1)+{\rm q}(p_2)+{\rm q}(p_3) \to {\rm q}(p_4)+{\rm q}
(p_5)+{\rm q}(p_6)$. The six momentum variables are assigned to quarks in six 
orderings.  
For instance, even though the momenta $p_1$, $p_2$ and $p_3$ may be assigned to
the initial quarks from left to right of a diagram, while $p_4$, $p_5$ and  
$p_6$ are assigned to the final quarks from left to right, the initial quarks  
may also take the order of ($p_1$, $p_3$, $p_2$) and correspondingly the final 
quarks have momenta in the order of ($p_4$, $p_6$, $p_5$). If these two  
orderings are denoted as  
\[  
\left ( \begin{array}{ccc}  
           p_4 & p_5 & p_6 \\ p_1 & p_2 & p_3   
           \end{array}   \right ),  
\left ( \begin{array}{ccc}  
           p_4 & p_6 & p_5 \\ p_1 & p_3 & p_2   
           \end{array}   \right ), 
\]   
the other four orderings are written as 
\[  
\left ( \begin{array}{ccc}  
           p_5 & p_4 & p_6 \\ p_2 & p_1 & p_3   
           \end{array}   \right ),  
\left ( \begin{array}{ccc}  
           p_5 & p_6 & p_4 \\ p_2 & p_3 & p_1   
           \end{array}   \right ), 
\left ( \begin{array}{ccc}  
           p_6 & p_4 & p_5 \\ p_3 & p_1 & p_2   
           \end{array}   \right ),  
\left ( \begin{array}{ccc}  
           p_6 & p_5 & p_4 \\ p_3 & p_2 & p_1   
           \end{array}   \right ). 
\]   
The six orderings applied to the six diagrams in Fig. 1 generate 36 different 
diagrams with the corresponding momentum labels. However, the six diagrams in  
Fig. 2 are unchanged upon using the six orderings. Therefore, 42 different  
diagrams with various momentum labels are needed to take into account the 
squared amplitudes. 
  
Nine Lorentz-invariant variables are defined as 
\begin{displaymath} 
s_{12}=(p_1+p_2)^2,~~~~~~~~s_{23}=(p_2+p_3)^2,~~~~~~~~s_{31}=(p_3+p_1)^2 
\end{displaymath} 
\begin{displaymath} 
u_{12} \equiv u_{15}=(p_1-p_5)^2,
~~~~~~~~~~~~~~~~~~u_{13} \equiv u_{16}=(p_1-p_6)^2 
\end{displaymath} 
\begin{displaymath} 
u_{21} \equiv u_{24}=(p_2-p_4)^2,
~~~~~~~~~~~~~~~~~~u_{23} \equiv u_{26}=(p_2-p_6)^2 
\end{displaymath} 
\begin{displaymath} 
u_{31} \equiv u_{34}=(p_3-p_4)^2,
~~~~~~~~~~~~~~~~~~u_{32} \equiv u_{35}=(p_3-p_5)^2 
\end{displaymath} 
The squared amplitudes and interference terms are expressed in terms of these  
independent variables for the three-quark to three-quark scatterings. Our  
Fortran code includes the squared amplitudes summed over spin and color states 
of the three final quarks and averaged over the spin and color states of  
the three initial quarks. All interference terms of amplitudes of different  
diagrams are also calculated. The two interference terms of a diagram in Fig. 1
 and a diagram in Fig. 2 cancel each other because the terms 
are pure imaginary. For example, 
\begin{equation} 
{\cal M}_{{\rm A}-45}^+ {\cal M}_{{\rm A}*4(56)}  
+ {\cal M}_{{\rm A}-45} {\cal M}_{{\rm A}*4(56)}^+ =0. 
\end{equation} 

We present the individually squared amplitudes and some
interference terms of the 
second class of diagrams in Fig. 2, which are not zero, as 
\begin{displaymath} 
\mid {\cal M}_{{\rm A}_*} \mid^2 = 3 \bar {A} / (w_1^2w_2^2w_3^2),
\end{displaymath} 
\begin{displaymath} 
{\cal M}_{{\rm A}_*} {\cal M}^+_{{\rm A}_{*4(56)}} =  \frac {3}{2} \bar {B}
/(u_{13}u_{21}u_{32}w_1w_2w_3),
\end{displaymath} 
\begin{displaymath} 
{\cal M}_{{\rm A}_*} {\cal M}^+_{{\rm A}_{*(45)6}} =  \frac {3}{2} \bar {C}
/(u_{12}u_{23}u_{31}w_1w_2w_3),
\end{displaymath} 
\begin{displaymath} 
\mid {\cal M}_{{\rm A}_{*45}} \mid^2 = 3 \bar {D} / 
(u_{12}^2u_{21}^2w_3^2),
\end{displaymath} 
\begin{displaymath} 
{\cal M}_{{\rm A}_{*45}} {\cal M}^+_{{\rm A}_{*56}} = - \frac {3}{2} \bar {E}
/(u_{12}u_{21}u_{23}u_{32}w_1w_3),
\end{displaymath} 
\begin{displaymath} 
{\cal M}_{{\rm A}_{*45}} {\cal M}^+_{{\rm A}_{*64}} = - \frac {3}{2} \bar {F}
/(u_{12}u_{21}u_{13}u_{31}w_2w_3),
\end{displaymath} 
\begin{displaymath} 
\mid {\cal M}_{{\rm A}_{*56}} \mid^2 = 3 \bar {G} / 
(u_{23}^2u_{32}^2w_1^2),
\end{displaymath} 
\begin{displaymath} 
{\cal M}_{{\rm A}_{*56}} {\cal M}^+_{{\rm A}_{*64}} = - \frac {3}{2} \bar {H}
/(u_{23}u_{32}u_{13}u_{31}w_1w_2),
\end{displaymath} 
\begin{displaymath} 
\mid {\cal M}_{{\rm A}_{*64}} \mid^2 = 3 \bar {I} / 
(u_{13}^2u_{31}^2w_2^2),
\end{displaymath} 
\begin{displaymath} 
\mid {\cal M}_{{\rm A}_{*4(56)}} \mid^2 = 3 \bar {J} / 
(u_{13}^2u_{21}^2u_{32}^2),
\end{displaymath} 
\begin{displaymath} 
{\cal M}_{{\rm A}_{*4(56)}} {\cal M}^+_{{\rm A}_{*(45)6}} = - \frac {3}{2} 
\bar {K} / (u_{13}u_{21}u_{32}u_{12}u_{23}u_{31}),
\end{displaymath} 
\begin{displaymath} 
\mid {\cal M}_{{\rm A}_{*(45)6}} \mid^2 = 3 \bar {L} / 
(u_{12}^2u_{23}^2u_{31}^2),
\end{displaymath} 
where 
$w_1=s_{12}+s_{31}+u_{12}+u_{13}$, $w_2=s_{12}+s_{23}+u_{21}+u_{23}$ and
$w_3=s_{31}+s_{23}+u_{31}+u_{32}$.
$\bar A$ is a function of the nine Lorentz-invariant variables,
$\bar {A}=\sum 
A(n_1,n_2,n_3,n_4,n_5,n_6,n_7,n_8,n_9)$
$\times s_{12}^{n_1} s_{23}^{n_2} s_{31}^{n_3} u_{12}^{n_4}
u_{13}^{n_5} u_{21}^{n_6} u_{23}^{n_7} u_{31}^{n_8} u_{32}^{n_9}$
where the sum is over $n_1$, $n_2$, $n_3$, $n_4$, $n_5$, $n_6$, $n_7$, $n_8$
and $n_9$, which run from 0 to 4 and are restricted to give
$n_1+n_2+n_3+n_4+n_5+n_6+n_7+n_8+n_9=4$. The non-zero A's are tabulated to 
form the column under $A$ in Table 1. 
$\bar {B}, \bar {C}, \bar {D}, \bar {E}, 
\bar {F}, \bar {G}, \bar {H}, \bar {I}, \bar {J}, \bar {K}$ and
$\bar {L}$ are similarly defined and
$B, C, D, E, F, G, H, I, J, K$ and $L$ can be found in Table 1. The above
interference terms give the other interference terms, for example,
${\cal M}^+_{{\rm A}_*} {\cal M}_{{\rm A}_{*4(56)}}=
{\cal M}_{{\rm A}_*} {\cal M}^+_{{\rm A}_{*4(56)}}$.

\vspace{0.5cm} 
\leftline{\bf 3. Thermalization of quark matter} 
\vspace{0.5cm} 
To study the thermalization of quark matter, these three-quark to three-quark  
scatterings are incorporated into a Boltzmann-type transport equation   
\begin{eqnarray} 
& &  
\frac {\partial f_1}{\partial t}  
+ \vec {\rm v}_1 \cdot \vec {\nabla}_{\vec r} f_1  
         \nonumber    \\ 
& & 
= -\frac {{\rm g}_{\rm Q}}{2E_1} \int \frac {d^3p_2}{(2\pi)^32E_2} 
\frac {d^3p_3}{(2\pi)^32E_3} \frac {d^3p_4}{(2\pi)^32E_4}   
(2\pi)^4 \delta^4(p_1+p_2-p_3-p_4) 
         \nonumber    \\ 
& & 
~~~ \times (\frac {1}{2} \mid {\cal M}_{{\rm uu} \to {\rm uu}} \mid^2  
+ \mid {\cal M}_{{\rm ud} \to {\rm ud}} \mid^2 ) 
[f_1f_2(1-f_3)(1-f_4)-f_3f_4(1-f_1)(1-f_2)] 
         \nonumber    \\ 
& & 
~~~ -\frac {{\rm g}_{\rm Q}^2}{2E_1}  
\int \frac {d^3p_2}{(2\pi)^32E_2} 
\frac {d^3p_3}{(2\pi)^32E_3} \frac {d^3p_4}{(2\pi)^32E_4}   
\frac {d^3p_5}{(2\pi)^32E_5} \frac {d^3p_6}{(2\pi)^32E_6}   
         \nonumber    \\ 
& & 
~~~ \times (2\pi)^4 \delta^4(p_1+p_2+p_3-p_4-p_5-p_6) 
         \nonumber    \\ 
& & 
~~~ \times [\frac {1}{12} \mid {\cal M}_{{\rm uuu} \to {\rm uuu}} \mid^2  
+\frac {1}{4} ( \mid {\cal M}_{{\rm uud} \to {\rm uud}} \mid^2  
              + \mid {\cal M}_{{\rm udu} \to {\rm udu}} \mid^2 )  
+\frac {1}{4} \mid {\cal M}_{{\rm udd} \to {\rm udd}} \mid^2]  
         \nonumber    \\ 
& & 
~~~ \times [f_1f_2f_3(1-f_4)(1-f_5)(1-f_6)-f_4f_5f_6(1-f_1)(1-f_2)(1-f_3)] 
         \nonumber    \\ 
\end{eqnarray} 
where the degeneracy factor ${\rm g}_{\rm Q}=6$ 
and the velocity of a massless quark ${\rm v}_1=1$.  
The distribution function $f_{\rm i}$ depends on  
the position $\vec {r}_{\rm i}$, the momentum $\vec {p}_{\rm i}$  
and the time $t$. The distribution function $f_1$ in this equation gives the 
up-quark distribution in quark matter. We assume that quark matter studied 
here has the same down-quark distribution as the up-quark distribution.  
The first term on the right-hand side of the above equation is for the  
2-quark to 2-quark scatterings. The second term is a new term representing 
the  3-quark to 3-quark scatterings and involving six quark distributions. 

The squared amplitude $\mid {\cal M}_{{\rm uuu} \to {\rm uuu}} \mid^2$ is for 
the scatterings of three identical quarks. It equals the sum of the  
individually squared amplitudes for the 42 diagrams discussed in the last 
section and the amplitudes corresponding to the interference terms between  
these diagrams. Both of the squared amplitudes, $\mid {\cal M}_{{\rm uud} \to  
{\rm uud}} \mid^2$ and $\mid {\cal M}_{{\rm udu} \to {\rm udu}} \mid^2$, are  
for the scatterings of two up-quarks and a down-quark. While the former  
receives contributions only from the diagrams ${\rm A}_{-}$, ${\rm A}_{-45}$,  
${\rm A}_{*}$ and ${\rm A}_{*45}$, the latter has contributions only from the  
diagrams ${\rm A}_{-}$, ${\rm A}_{-64}$, ${\rm A}_{*}$ and ${\rm A}_{*64}$.  
The squared amplitude $\mid {\cal M}_{{\rm udd} \to {\rm udd}} \mid^2$  
is for the scatterings of one up-quark and two down-quarks.  
Only the diagrams ${\rm A}_{-}$, ${\rm A}_{-56}$, ${\rm A}_{*}$ and  
${\rm A}_{*56}$ contribute to the squared amplitude. 

\vspace{0.5cm} 
\leftline{\bf 4. Numerical results} 
\vspace{0.5cm} 
Anisotropic parton momentum distributions are formed in initial Au-Au  
collisions. Such anisotropy can be eliminated by elastic scatterings among  
partons. Starting from the time $t_{\rm ini},$ when anisotropic quark matter  
is formed and ending at the time $t_{\rm iso},$ when local momentum isotropy is
established, the transport equation is solved. The squared amplitudes are  
calculated at $\alpha_{\rm s}=0.6$. In the calculations, a  
divergence of the squared amplitudes is encountered. The divergence is removed 
while gluon propagators are regularized by a screening mass which is evaluated
from the distribution function by a formula used in Refs. [17-19]. We 
arbitrarily add 100 MeV to or subtract 100 MeV from the screening mass, 
a change of thermalization time by about 40\% is obtained. Any two runs of
Fortran codes for solving the transport equation do not give rise to a 
difference of screening masses more than 100 MeV. Therefore, the uncertainty in
the thermalization time due to the screening mass is less than 40\%. 

Since the initial gluon rapidity density  
accounting for the RHIC data [12,13] is about 1000, 
larger than 200 given by HIJING [22], 
the initial quark rapidity density must also be similarly underestimated
by HIJING simulation. 
For a central Au-Au collision at $\sqrt {s_{NN}}=200$ GeV, 
an initial quark momentum distribution can be, for example, four times [23] 
as large as that obtained from HIJING simulation [24],
\begin{equation} 
f(\vec {p},t_{\rm ini})=\frac {7.12\times 10^5 (2\pi)^{1.5}} 
{\pi R_A^2 Y(\mid \vec {p} \mid/\cosh ({\rm y})+0.3)} 
{\rm e}^{-\mid \vec {p} \mid/(0.9\cosh ({\rm y}))-(\mid \vec {p} \mid  
\tanh ({\rm y}))^2/8} \bar {\theta} (Y^2-{\rm y}^2) 
\end{equation} 
where $\mid \vec {p} \mid$ is in GeV, 
the nuclear radius $R_A=6.4$ fm, $t_{\rm ini}$ = 0.2 fm/$c$ and  
the rapidity y is less than $Y$ which approximately equals 5.  
$\bar {\theta} (x)$ is the step function which is zero for $x<0$ 
or 1 for $x \ge 0$. Anisotropy of the distribution is shown in Fig. 3 
by the differences of dotted, dashed and dot-dashed curves, which represent 
the quark momentum distributions at the three angles  
$\theta = 0^{\rm o}, 45^{\rm o}, 90^{\rm o}$ relative to one incoming gold beam
direction. The quark momentum distribution in the longitudinal direction  
is considerably larger than one in the transverse direction when momentum  
departs from 0. 

Numerical simulations of the $2 \to 2$ and $3 \to 3$ scatterings in Fortran 
codes depend on quark positions. Starting from the cross section 
for the 2-quark to 2-quark scattering $\sigma_{2 \to 2}$ which is 
calculated from the squared amplitudes in Eqs. (1)-(2), a scattering 
of two quarks is considered to occur when the closest distance of the two 
quarks is less than an interaction range of $\sqrt {\sigma_{2 \to 2}/\pi}$.
Regarding to a three-quark scattering, the three quarks must be in a sphere
which center is at the center-of-mass of the three quarks and which radius 
$r_{\rm hs}$ is determined by [11]  
\begin{eqnarray}
\pi r_{\rm hs}^2       
& = & \frac {1}{3!} \int \frac {d^3p_4}{(2\pi)^32E_4}  
\frac {d^3p_5}{(2\pi)^32E_5} \frac {d^3p_6}{(2\pi)^32E_6}  
         \nonumber    \\
& &
\times (2\pi)^4 \delta^4(p_1+p_2+p_3-p_4-p_5-p_6)
\mid {\cal M}_{3 \to 3} \mid^2 
\end{eqnarray}
where $\mid {\cal M}_{3 \to 3} \mid^2$ is the squared amplitude for the
3-quark to 3-quark scattering. One consequence of such numerical simulations 
is to unfortunately break 
the locality of the transport equation and thus Lorentz covariance.
Nevertheless,
the particle  subdivision technique [20,21] can restore the Lorentz covariance
[11] by means of the transformation        
\begin{displaymath}
f \to f'=\ell f,
\end{displaymath}
\begin{displaymath}
\mid {\cal M}_{2 \to 2} \mid^2 \to  
\mid {\cal M}_{2 \to 2}' \mid^2 = \mid {\cal M}_{2 \to 2} \mid^2 /\ell,
~~~\mid {\cal M}_{3 \to 3} \mid^2 \to  
\mid {\cal M}_{3 \to 3}' \mid^2 = \mid {\cal M}_{3 \to 3} \mid^2 /\ell^2,
\end{displaymath}
in combination with the replacement
\begin{displaymath}
[f_1f_2(1-f_3)(1-f_4)-f_3f_4(1-f_1)(1-f_2)] \to f_1f_2-f_3f_4 
\end{displaymath}
\begin{displaymath}
[f_1f_2f_3(1-f_4)(1-f_5)(1-f_6)-f_4f_5f_6(1-f_1)(1-f_2)(1-f_3)]
\to f_1f_2f_3-f_4f_5f_6 
\end{displaymath}
In principle, the Lorentz covariance is recovered while $\ell$ goes to 
infinity. The practical value of $\ell$ in codes equals 40. Six hundred and
sixty-six quarks are generated from the initial distribution (5) and the 
average of 20 runs of Fortran codes with different random number sets 
is taken as a solution of Eq. (4). 
  
The solution of the transport equation at $t_{\rm iso}=2$ fm/$c$ 
is shown by the dotted, dashed and dot-dashed curves in Fig. 4.    
The curves overlap and can thus be fitted to the J$\rm \ddot u$ttner  
distribution, 
\begin{equation} 
f(\vec {p},t_{\rm iso})=\frac {\lambda}{{\rm e}^{\mid \vec {p} \mid/T}-\lambda}
\end{equation} 
where the temperature of quark matter $T=0.59$ GeV and fugacity  
$\lambda=0.04$. We get a thermalization time of $t_{\rm iso}-t_{\rm ini}=1.8$ 
fm/$c$. For strongly interacting quark matter, the present value of the
coupling constant $\alpha_{\rm s}=0.6$ is allowed [25]. For weakly interacting
quark matter, $\alpha_{\rm s}$ needs to be reduced.
If the initial quark distribution and/or the coupling constant
becomes smaller, the thermalization time of quark matter gets longer than
the order of 1.8 fm/$c$.

\vspace{0.5cm} 
\leftline{\bf 5. Summary} 
\vspace{0.5cm} 
We have studied the thermalization of quark matter including three-quark  
elastic scatterings in the transport equation. Squared amplitudes for   
triple-quark scatterings are derived at the tree level in QCD. Given an  
anisotropic quark distribution at the formation of quark matter, momentum  
isotropy can be shown by the transport equation solution. The history of quark
matter's evolving into a thermal state can be described by the transport 
equation. The triple-quark scatterings give a variation of quark 
distribution function comparable to the one that resulted from the quark-quark
scatterings. Rapid thermalization can not be established by the  
triple-quark scatterings from such an anisotropic quark distribution which  
can appropriately describe quark matter initially created in central Au-Au 
collisions at the highest RHIC energy. Therefore, we also conclude that rapid
thermalization of quark matter must depend on the gluon-gluon-quark, 
gluon-quark-quark and gluon-antiquark-quark scatterings which involve higher
gluon number density than the quark number density
and have larger squared amplitudes than the
quark-quark-quark scatterings. Such very complicated works including the
quark scatterings with gluons and/or antiquarks remain to be done.  

\vspace{0.5cm} 
\leftline{\bf Acknowledgements} 
\vspace{0.5cm} 
This work was supported in part by National Natural Science Foundation of China
under Grant No. 10135030, in part by Shanghai Education Committee Research  
Fund and in part by the CAS Knowledge Innovation Project No. KJCX2-SW-N02. 

\newpage 
\leftline{\bf References} 
\vskip 14pt 
\leftline{[1]U. Heinz, P. F. Kolb, in: R. Bellwied, J. Harris, W. Bauer 
(Eds.), 
 Proc. of the 18th}\leftline{~~~Winter Workshop on Nuclear Dynamics, EP  
Systema, Debrecen, Hungary, 2002.} 
\leftline{[2]E. V. Shuryak, Nucl. Phys. {\bf A715}(2003)289c.}   
\leftline{[3]T. Hirano, Phys. Rev. {\bf C65}(2001)011901.} 
\leftline{[4]P. Huovinen, Nucl. Phys. {\bf A715}(2003)299c.} 
\leftline{[5]K. Morita, S. Muroya, C. Nonaka, T. Hirano, Phys. Rev. 
{\bf C66}(2002)054904.} 
\leftline{[6]D. Teaney, J. Lauret, E. V. Shuryak, nucl-th/0110037.} 
\leftline{[7]K. J. Eskola, H. Niemi, P. V. Ruuskanen, S. S.  
R$\rm \ddot a$s$\rm \ddot a$nen, Phys. Lett. {\bf B566}(2003)187;} 
\leftline{~~~K. J. Eskola, H. Niemi, P. V. Ruuskanen, S. S.  
R$\rm \ddot a$s$\rm \ddot a$nen, Nucl. Phys. {\bf A715}(2003)561c.} 
\leftline{[8]K. H. Ackermann, et al., STAR Collaboration, Phys. Rev. Lett.  
{\bf 86}(2001)402;} 
\leftline{~~~R. J. Snellings, et al., for the STAR Collaboration, Nucl. Phys.  
{\bf A698}(2002)193c;}  
\leftline{~~~C. Adler, et al., STAR Collaboration, Phys. Rev. Lett.  
{\bf 87}(2001)182301;} 
\leftline{~~~C. Adler, et al., STAR Collaboration, Phys. Rev.  
{\bf C66}(2002)034904;}  
\leftline{~~~C. Adler, et al., STAR Collaboration, Phys. Rev. Lett.  
{\bf 90}(2003)032301.} 
\leftline{[9]R. A. Lacey, et al., for the PHENIX Collaboration, Nucl. Phys.  
{\bf A698}(2002)559c;} 
\leftline{~~~S. S. Adler, et al., PHENIX Collaboration, Phys. Rev. Lett.  
{\bf 91}(2003)182301;}  
\leftline{~~~S. S. Adler, et al., PHENIX Collaboration, nucl-ex/0411040.}  
\leftline{[10]K. Geiger, Phys. Rev. {\bf D46}(1992)4965;} 
\leftline{~~~~K. Geiger, Phys. Rev. {\bf D46}(1992)4986.} 
\leftline{[11]X.-M. Xu, Y. Sun, A.-Q. Chen, L. Zheng, Nucl. Phys.  
{\bf A744}(2004)347.} 
\leftline{[12]K. J. Eskola, K. Kajantie, K. Tuominen, Phys. Lett. {\bf B497} 
(2001)39.} 
\leftline{[13]M. Gyulassy, P. L$\rm \acute e$vai, I. Vitev, Nucl. Phys.  
{\bf B594}(2001)371;}  
\leftline{~~~~M. Gyulassy, I. Vitev, X.-N. Wang, P. Huovinen,  
Phys. Lett. {\bf B526}(2001)301.}  
\leftline{[14]F. Cooper, E. Mottola, G. C. Nayak, Phys. Lett. {\bf B555}(2003) 
181.} 
\leftline{[15]R. Cutler, D. Sivers, Phys. Rev. {\bf D17}(1978)196.} 
\leftline{[16]B. L. Combridge, J. Kripfganz, J. Ranft, Phys. Lett.  
{\bf B70}(1977)234.} 
\leftline{[17]T. S. Bir$\rm \acute o$, B. M$\rm \ddot u$ller, 
X.-N. Wang, Phys. Lett. {\bf B283}(1992)171.} 
\leftline{[18]K. J. Eskola, B. M$\rm \ddot u$ller, 
X.-N. Wang, Phys. Lett. {\bf B374}(1996)20.} 
\leftline{[19]S.A. Bass, B. M$\rm \ddot u$ller,
D. K. Srivastava, Phys. Lett. {\bf B551}(2003)277.} 
\leftline{[20]B. Zhang, M. Gyulassy, Y. Pang, Phys. Rev. {\bf C58}(1998)1175.} 
\leftline{[21]D. Moln$\rm \acute a$r, M. Gyulassy, Nucl. Phys. 
{\bf A697}(2002)495.}
\leftline{[22]X.-N. Wang, M. Gyulassy, Phys. Rev. {\bf D44}(1991)3501;} 
\leftline{~~~~X.-N. Wang, M. Gyulassy, Comput. Phys. Commun. {\bf 83}(1994)307;}\leftline{~~~~X.-N. Wang, Phys. Rep. {\bf 280}(1997)287.} 
\leftline{[23]X.-M. Xu, D. Kharzeev, H. Satz, X.-N. Wang, Phys. Rev. {\bf C53}
(1996)3051.}
\leftline{[24]P. L$\rm \acute e$vai, B. M$\rm \ddot u$ller, X.-N. Wang, Phys.  
Rev. {\bf C51}(1995)3326.} 
\leftline{[25]L.-W. Chen, C.M. Ko, Z.-W. Lin, Phys. Rev. 
{\bf C69}(2004)031901.}

\newpage 
\begin{figure}[t] 
  \begin{center} 
    \leavevmode 
    \parbox[t]{\textwidth} 
            {\psfig{file=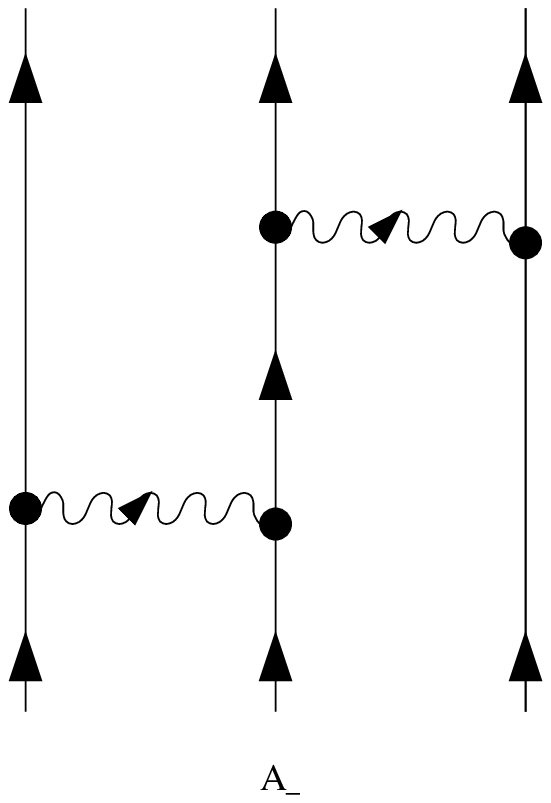,width=42mm,height=65mm,angle=0}} 
            \hspace{1.2cm} 
    \parbox[t]{\textwidth} 
            {\psfig{file=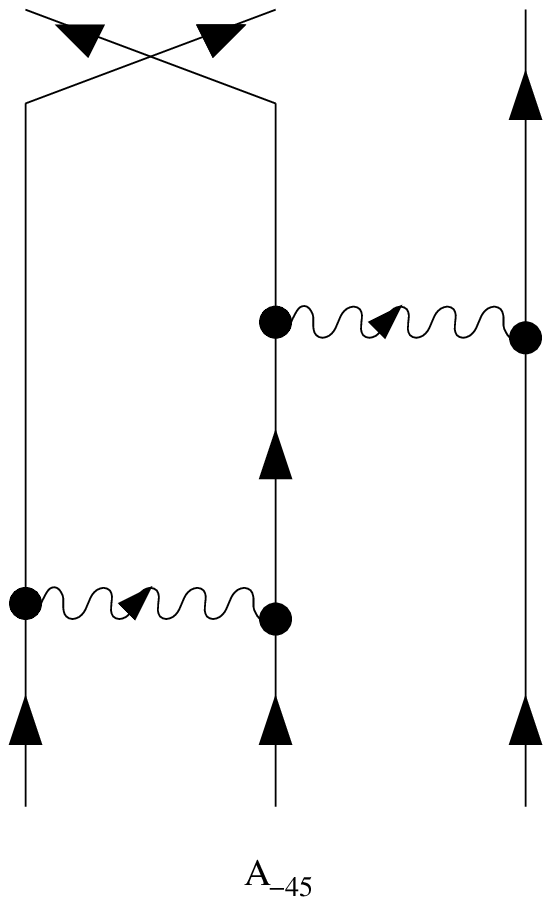,width=42mm,height=65mm,angle=0}} 
            \hspace{1.2cm} 
    \parbox[t]{\textwidth} 
            {\psfig{file=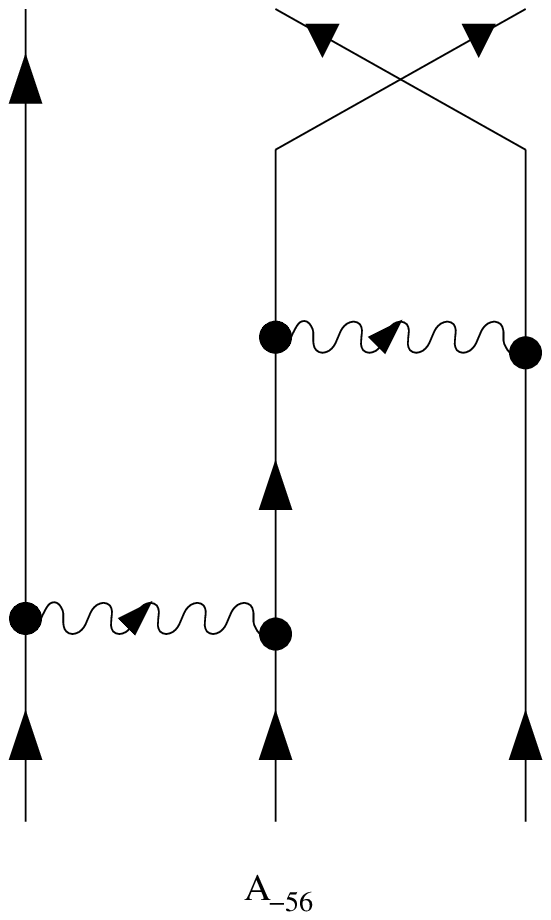,width=42mm,height=65mm,angle=0}} 
            \vskip 26pt 
    \parbox[t]{\textwidth} 
            {\psfig{file=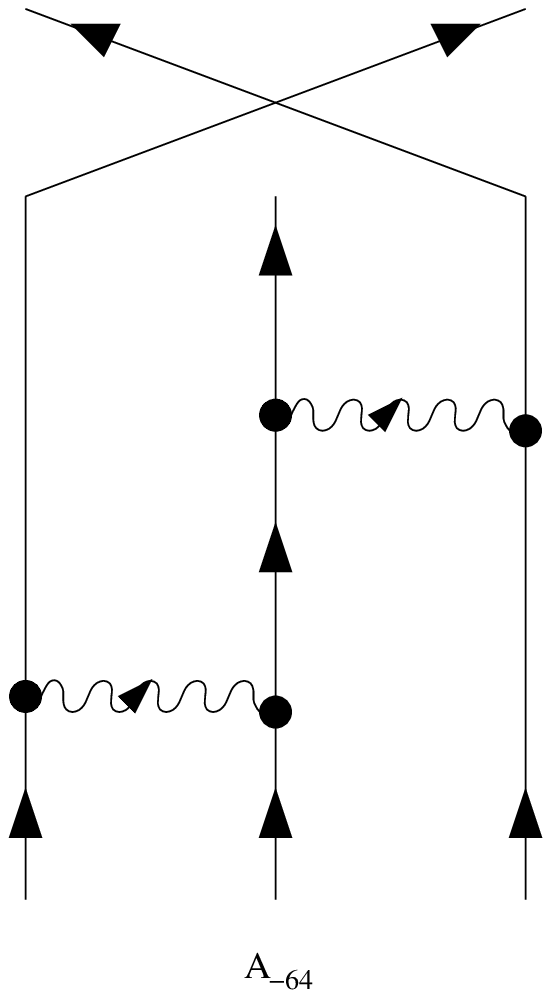,width=42mm,height=65mm,angle=0}} 
            \hspace{1.2cm}  
    \parbox[t]{\textwidth} 
            {\psfig{file=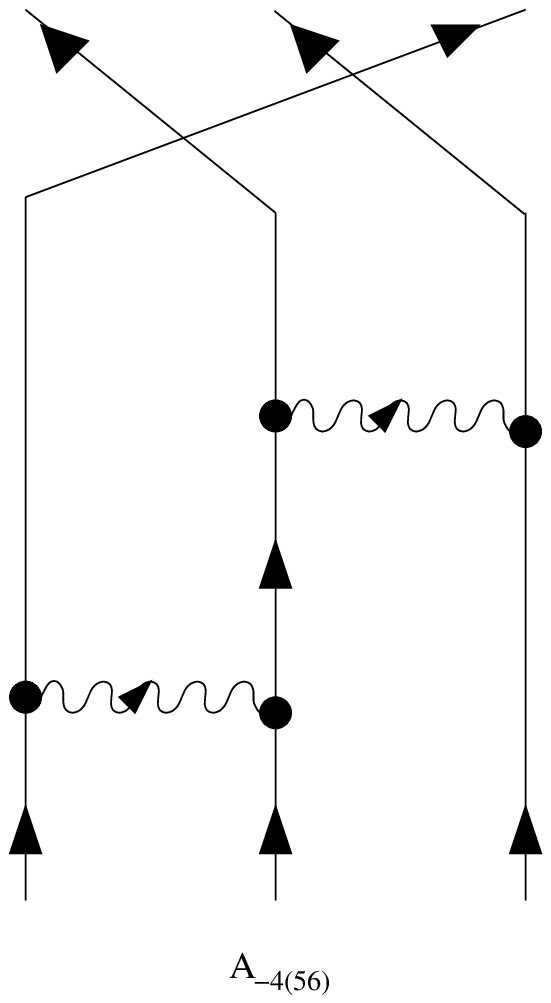,width=42mm,height=65mm,angle=0}} 
            \hspace{1.2cm}  
    \parbox[t]{\textwidth} 
            {\psfig{file=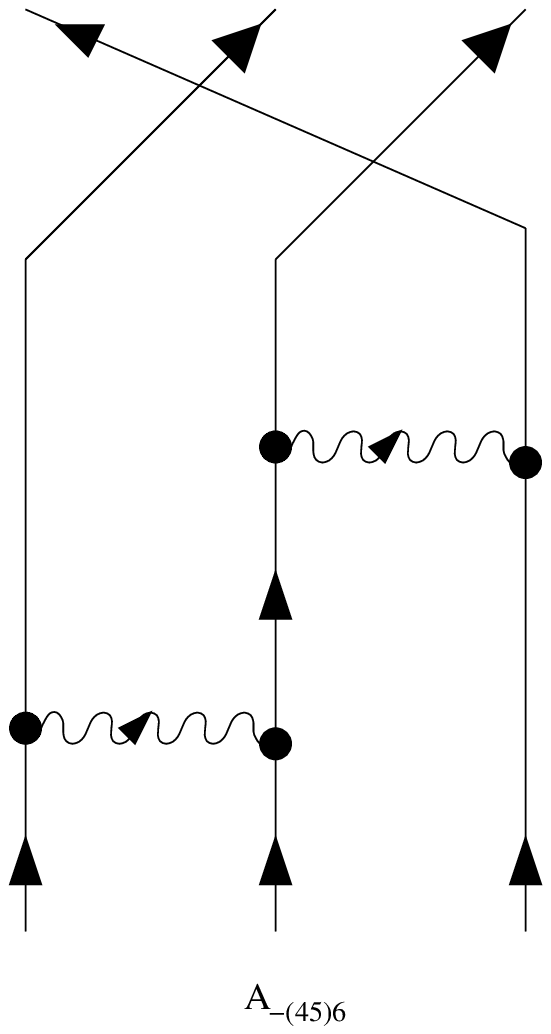,width=42mm,height=65mm,angle=0}} 
  \end{center} 
\caption{Two-gluon-exchange induced scatterings of three quarks.}  
\label{fig1} 
\end{figure} 

\newpage 
\begin{figure}[t] 
  \begin{center} 
    \parbox[t]{\textwidth} 
            {\psfig{file=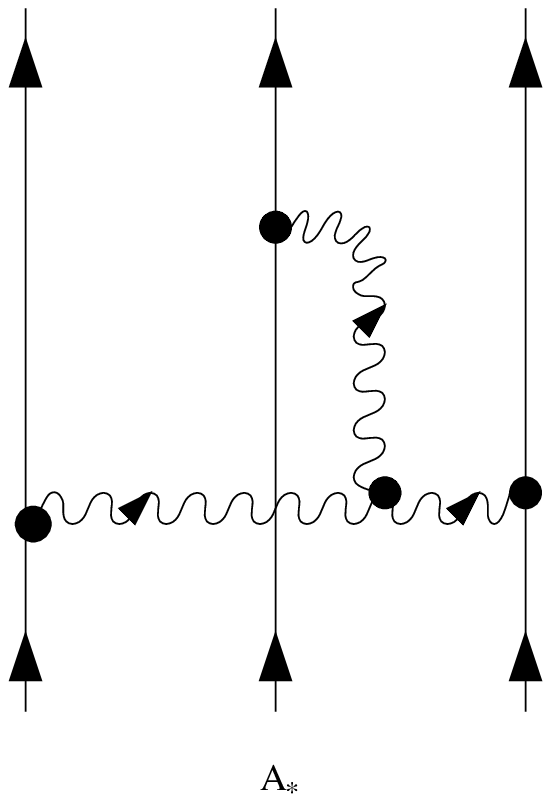,width=42mm,height=65mm,angle=0}} 
            \hspace{1.2cm}    
    \parbox[t]{\textwidth} 
            {\psfig{file=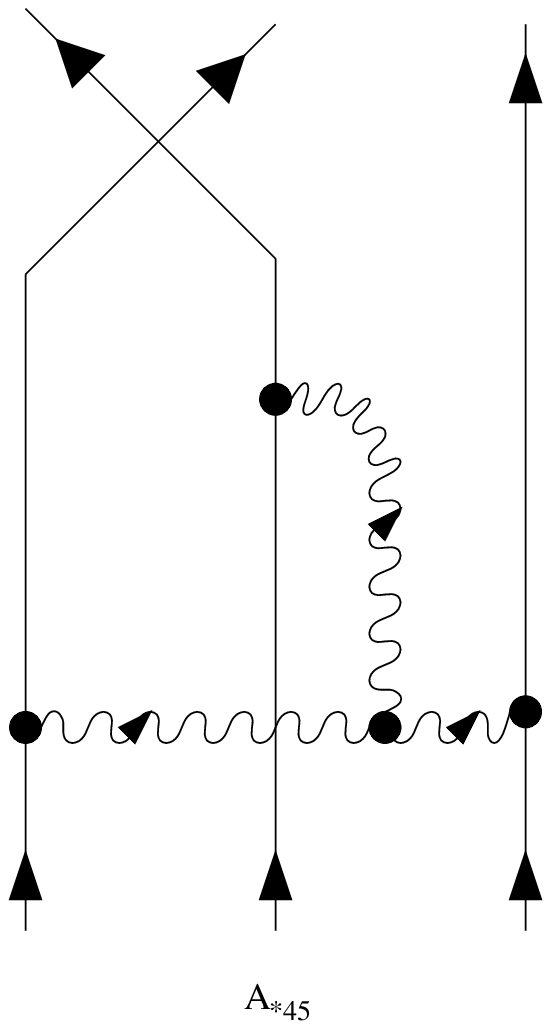,width=42mm,height=65mm,angle=0}} 
            \hspace{1.2cm}    
    \parbox[t]{\textwidth} 
            {\psfig{file=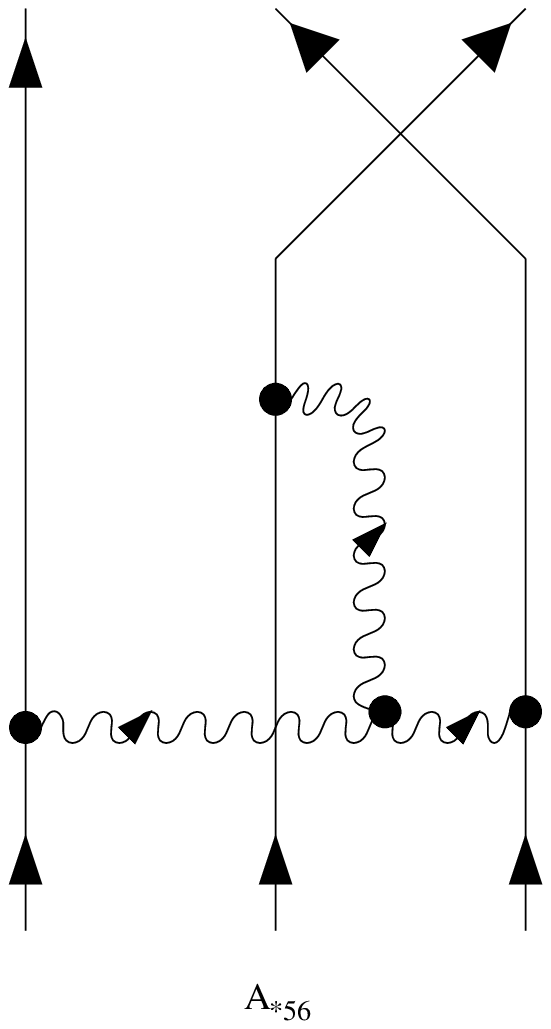,width=42mm,height=65mm,angle=0}} 
            \vskip 26pt 
    \parbox[t]{\textwidth} 
            {\psfig{file=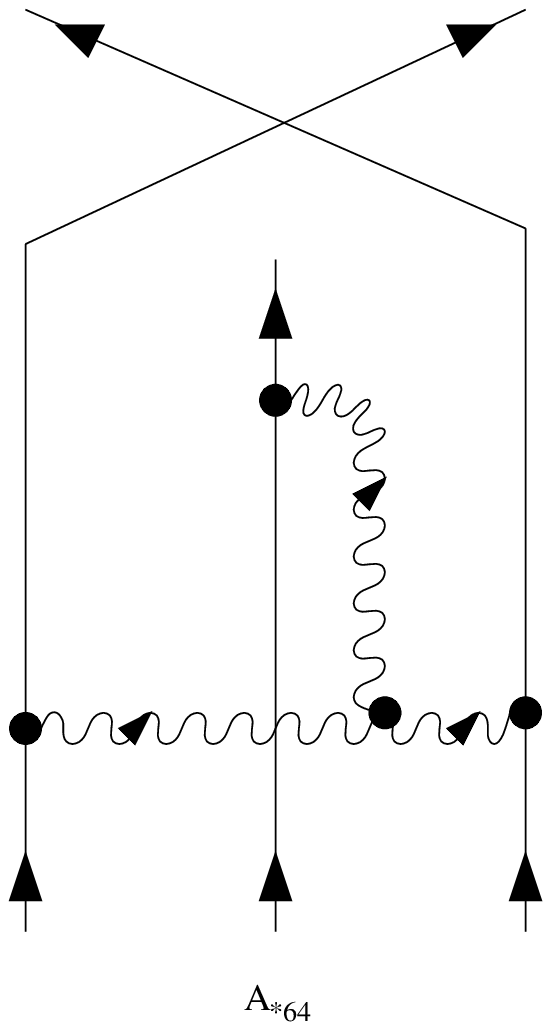,width=42mm,height=65mm,angle=0}} 
            \hspace{1.2cm}    
    \parbox[t]{\textwidth} 
            {\psfig{file=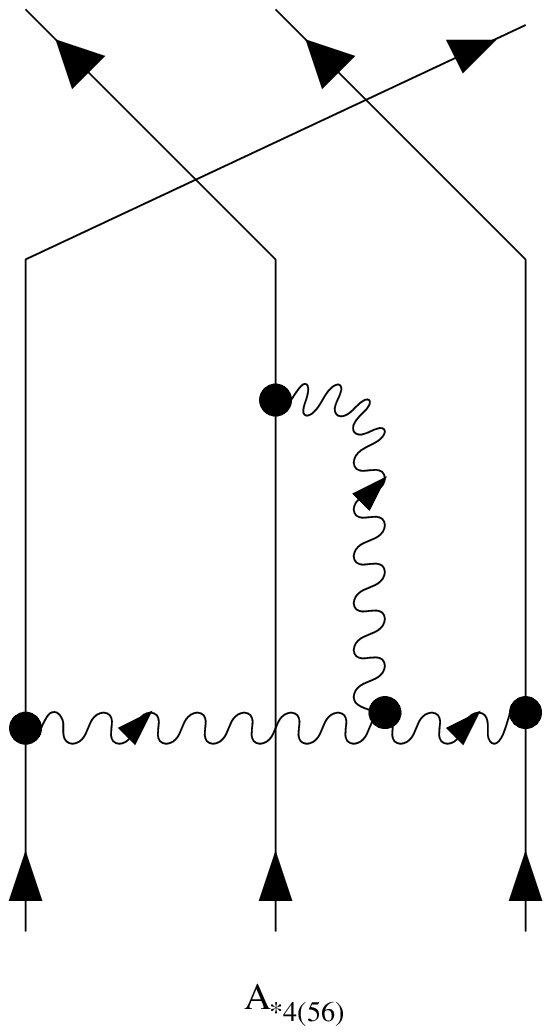,width=42mm,height=65mm,angle=0}} 
            \hspace{1.2cm}  
    \parbox[t]{\textwidth} 
            {\psfig{file=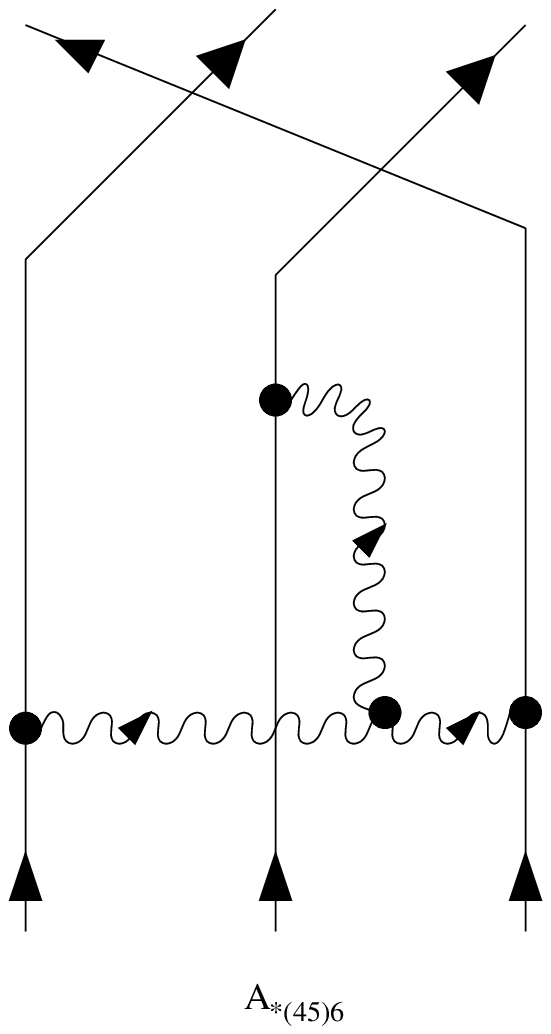,width=42mm,height=65mm,angle=0}} 
  \end{center} 
\caption{Triple-gluon coupling in scatterings of three quarks.} 
\label{fig2} 
\end{figure}

\newpage 
\begin{figure}[t] 
  \begin{center} 
    \parbox{\textwidth} 
           {\psfig{file=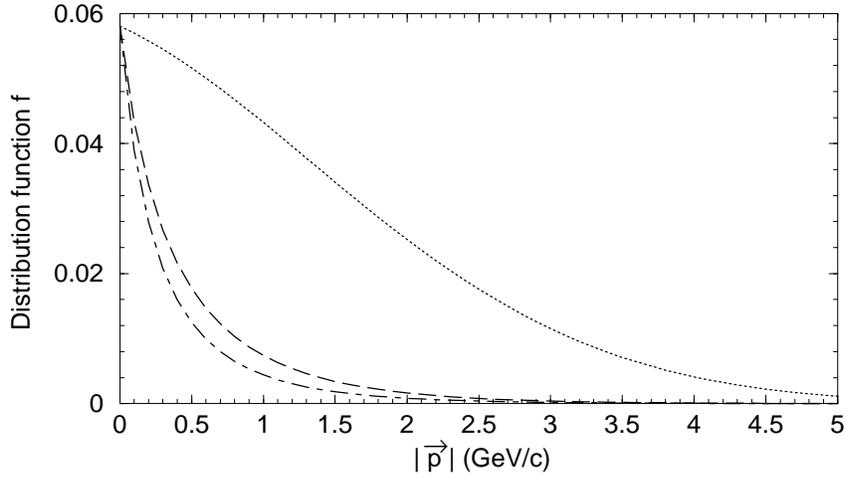,width=0.7\textwidth,angle=0}} 
  \end{center} 
\caption{Quark distribution functions versus momentum in different 
directions while anisotropic quark matter is produced in the  
initial Au-Au collision. The dotted, dashed and dot-dashed curves correspond  
to the angles $\theta =0^{\rm o}, 45^{\rm o}, 90^{\rm o}$, respectively.} 
\label{fig3} 
\end{figure} 

\newpage 
\begin{figure}[t] 
  \begin{center} 
    \parbox{\textwidth} 
           {\psfig{file=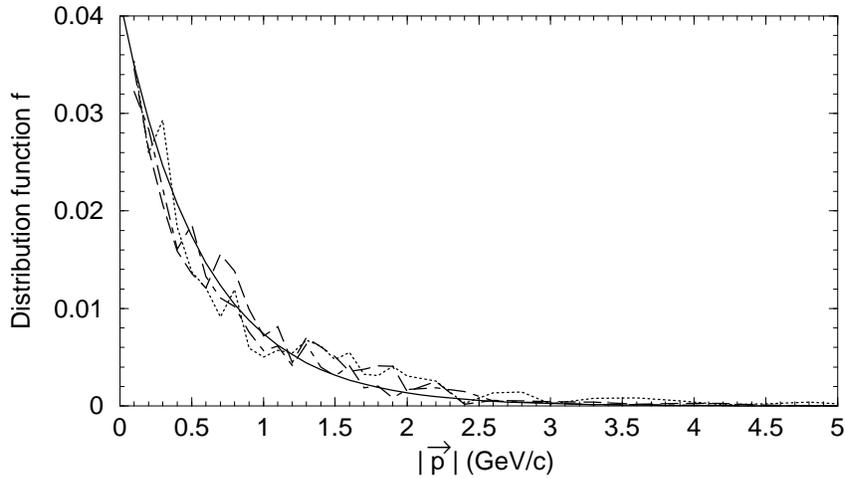,width=0.7\textwidth,angle=0}} 
  \end{center} 
\caption{Quark distribution functions versus momentum in different 
directions while  quark matter just arrives at thermal equilibrium. 
The dotted, dashed and dot-dashed curves correspond to the angles   
$\theta =0^{\rm o}, 45^{\rm o}, 90^{\rm o}$, respectively.  
The solid curve represents the thermal distribution function.} 
\label{fig4} 
\end{figure} 

\begin{table}
\caption{For the entries 
we use the symbols $(a_1,\cdot \cdot \cdot, a_{23})$
=(1024, 512, 2560, 1536, 2048, 3072, 128, 256, 384, 768, 640,
896, 1152, 1280, 2304, 1792, 3328, 4096, 64, 448, 320, 192, 832)/3456.} 
\begin{center}
\begin{tabular}{lllllllllllll}
\hline\hline
& $A$ & $B$ & $C$ & $D$  & $E$  & $F$ 
& $G$ & $H$ & $I$ & $J$ & $K$ & $L$ \\
\hline
$u_{13}u_{23}u_{32}^2$ 
& 0 & 0 & 0 & -$a_{2}$ 
& 0 & 0 & -$a_{1}$ & 0 
& -$a_{1}$ & -$a_{10}$ & 0 & 0 \\
$u_{13}u_{23}u_{31}u_{32}$ 
& -$a_{1}$ & 0 & 0 & -$a_{1}$ 
& 0 & 0 & -$a_{4}$ & 0 
& -$a_{4}$ & -$a_{10}$ & 0 & -$a_{8}$ \\
$u_{13}u_{23}u_{31}^2$ 
& 0 & 0 & 0 & -$a_{2}$ 
& 0 & 0 & -$a_{1}$ & 0 
& -$a_{1}$ & -$a_{2}$ & 0 & -$a_{8}$ \\
$u_{13}u_{23}^2u_{32}$ 
& $a_{1}$ & 0 & 0 & 0 
& 0 & 0 & 0 & 0 
& $a_{2}$ & $a_{1}$ & 0 & $a_{8}$ \\
$u_{13}u_{23}^2u_{31}$ 
& $a_{1}$ & 0 & 0 & 0 
& 0 & 0 & $a_{2}$ & 0 
& $a_{1}$ & $a_{2}$ & 0 & $a_{8}$ \\
$u_{13}u_{21}u_{32}^2$ 
& $a_{2}$ & 0 & 0 & 0 
& 0 & 0 & 0 & 0 
& -$a_{1}$ & 0 & 0 & 0 \\
$u_{13}u_{21}u_{31}u_{32}$ 
& $a_{2}$ & 0 & 0 & $a_{2}$ 
& 0 & 0 & -$a_{2}$ & 0 
& -$a_{1}$ & $a_{8}$ & 0 & -$a_{8}$ \\
$u_{13}u_{21}u_{31}^2$ 
& $a_{1}$ & 0 & 0 & $a_{2}$ 
& 0 & 0 & 0 & 0 
& 0 & $a_{2}$ & 0 & 0 \\
$u_{13}u_{21}u_{23}u_{32}$ 
& $a_{2}$ & 0 & 0 & -$a_{2}$ 
& 0 & 0 & -$a_{1}$ & 0 
& $a_{2}$ & $a_{10}$ & 0 & $a_{8}$ \\
$u_{13}u_{21}u_{23}u_{31}$ 
& $a_{1}$ & 0 & 0 & -$a_{2}$ 
& 0 & 0 & -$a_{2}$ & 0 
& $a_{1}$ & 0 & 0 & 0 \\
$u_{13}u_{21}^2u_{32}$ 
& $a_{2}$ & 0 & 0 & 0 
& 0 & 0 & -$a_{1}$ & 0 
& 0 & $a_{2}$ & 0 & 0 \\
$u_{13}u_{21}^2u_{31}$ 
& $a_{1}$ & 0 & 0 & $a_{2}$ 
& 0 & 0 & 0 & 0
& $a_{1}$ & $a_{2}$ & 0 & 0 \\
$u_{13}^2u_{23}u_{32}$ 
& $a_{1}$ & 0 & 0 & 0 
& 0 & 0 & $a_{1}$ & 0 
& $a_{2}$ & $a_{8}$ & 0 & -$a_{2}$ \\
$u_{13}^2u_{23}u_{31}$ 
& $a_{1}$ & 0 & 0 & 0 
& 0 & 0 & $a_{2}$ & 0 
& 0 & $a_{1}$ & 0 & $a_{2}$ \\
$u_{13}^2u_{21}u_{32}$ 
& $a_{2}$ & 0 & 0 & -$a_{1}$ 
& 0 & 0 & 0 & 0 
& 0 & $a_{2}$ & 0 & -$a_{8}$ \\
$u_{13}^2u_{21}u_{31}$ 
& 0 & 0 & 0 & -$a_{1}$ 
& 0 & 0 & -$a_{2}$ & 0 
& -$a_{1}$ & 0 & 0 & $a_{8}$ \\
$u_{12}u_{23}u_{32}^2$ 
& $a_{1}$ & 0 & 0 & $a_{2}$ 
& 0 & 0 & 0 & 0 
& 0 & $a_{8}$ & 0 & $a_{1}$ \\
$u_{12}u_{23}u_{31}u_{32}$ 
& $a_{2}$ & 0 & 0 & $a_{2}$ 
& 0 & 0 & -$a_{1}$ & 0 
& -$a_{2}$ & $a_{8}$ & 0 & $a_{10}$ \\
$u_{12}u_{23}u_{31}^2$ 
& $a_{2}$ & 0 & 0 & 0 
& 0 & 0 & -$a_{1}$ & 0 
& 0 & 0 & 0 & $a_{2}$ \\
$u_{12}u_{23}^2u_{32}$ 
& 0 & 0 & 0 & -$a_{1}$ 
& 0 & 0 & -$a_{1}$ & 0 
& -$a_{2}$ & 0 & 0 & -$a_{10}$ \\
$u_{12}u_{23}^2u_{31}$ 
& $a_{2}$ & 0 & 0 & -$a_{1}$ 
& 0 & 0 & 0 & 0 
& 0 & 0 & 0 & 0 \\
$u_{12}u_{21}u_{32}^2$ 
& $a_{1}$ & 0 & 0 & $a_{1}$ 
& 0 & 0 & $a_{2}$ & 0 
& 0 & $a_{8}$ & 0 & $a_{2}$ \\
$u_{12}u_{21}u_{31}u_{32}$ 
& $a_{1}$ & 0 & 0 & $a_{1}$ 
& 0 & 0 & -$a_{2}$ & 0 
& -$a_{2}$ & 0 & 0 & 0 \\
$u_{12}u_{21}u_{31}^2$ 
& $a_{1}$ & 0 & 0 & $a_{1}$ 
& 0 & 0 & 0 & 0 
& $a_{2}$ & 0 & 0 & $a_{2}$ \\
$u_{12}u_{21}u_{23}u_{32}$ 
& -$a_{1}$ & 0 & 0 & -$a_{4}$ 
& 0 & 0 & -$a_{4}$ & 0 
& -$a_{1}$ & -$a_{8}$ & 0 & -$a_{10}$ \\
$u_{12}u_{21}u_{23}u_{31}$ 
& $a_{2}$ & 0 & 0 & -$a_{1}$ 
& 0 & 0 & -$a_{2}$ & 0 
& $a_{2}$ & -$a_{8}$ & 0 & $a_{8}$ \\
$u_{12}u_{21}^2u_{32}$ 
& 0 & 0 & 0 & -$a_{1}$ 
& 0 & 0 & -$a_{1}$ & 0 
& -$a_{2}$ & -$a_{8}$ & 0 & -$a_{2}$ \\
$u_{12}u_{21}^2u_{31}$ 
& $a_{1}$ & 0 & 0 & 0 
& 0 & 0 & 0 & 0 
& $a_{2}$ & 0 & 0 & $a_{2}$ \\
$u_{12}u_{13}u_{23}u_{32}$ 
& $a_{1}$ & 0 & 0 & -$a_{2}$ 
& 0 & 0 & $a_{1}$ & 0 
& -$a_{2}$ & -$a_{14}$ & 0 & -$a_{14}$ \\
$u_{12}u_{13}u_{23}u_{31}$ 
& $a_{2}$ & 0 & 0 & -$a_{2}$ 
& 0 & 0 & $a_{2}$ & 0 
& -$a_{1}$ & -$a_{8}$ & 0 & -$a_{8}$ \\
$u_{12}u_{13}u_{21}u_{32}$ 
& $a_{2}$ & 0 & 0 & -$a_{1}$ 
& 0 & 0 & $a_{2}$ & 0 
& -$a_{2}$ & -$a_{8}$ & 0 & -$a_{8}$ \\
$u_{12}u_{13}u_{21}u_{31}$ 
& -$a_{1}$ & 0 & 0 & -$a_{4}$ 
& 0 & 0 & -$a_{1}$ & 0 
& -$a_{4}$ &  -$a_{2}$ & 0 & -$a_{2}$ \\
$u_{12}^2u_{23}u_{32}$ 
& $a_{1}$ & 0 & 0 & $a_{2}$ 
& 0 & 0 & $a_{1}$ & 0 
& 0 & -$a_{2}$ & 0 & $a_{8}$ \\
$u_{12}^2u_{23}u_{31}$ 
& $a_{2}$ & 0 & 0 & 0 
& 0 & 0 & 0 & 0 
& -$a_{1}$ & -$a_{8}$ & 0 & $a_{2}$ \\
$u_{12}^2u_{21}u_{32}$ 
& $a_{1}$ & 0 & 0 & 0 
& 0 & 0 & $a_{2}$ & 0 
& 0 & $a_{2}$ & 0 & $a_{1}$ \\
$u_{12}^2u_{21}u_{31}$ 
& 0 & 0 & 0 & -$a_{1}$ 
& 0 & 0 & -$a_{2}$ & 0 
& -$a_{1}$ & $a_{8}$ & 0 & 0 \\

\hline\hline
\end{tabular}
\end{center}
\end{table}

\begin{table}
\centerline{Table 1. Continued.}
\begin{center}
\begin{tabular}{lllllllllllll}
\hline\hline
& $A$ & $B$ & $C$ & $D$  & $E$  & $F$ 
& $G$ & $H$ & $I$ & $J$ & $K$ & $L$ \\
\hline
$s_{31}u_{23}u_{32}^2$ 
& $a_{2}$ & 0 & 0 & $a_{2}$ 
& 0 & 0 & -$a_{2}$ & 0 
& 0 & -$a_{8}$ & 0 & 0 \\
$s_{31}u_{23}u_{31}u_{32}$ 
& 0 & 0 & 0 & 0
& 0 & 0 & -$a_{1}$ & 0 
& -$a_{1}$ & -$a_{8}$ & 0 & -$a_{2}$ \\
$s_{31}u_{23}u_{31}^2$ 
& $a_{2}$ & 0 & 0 & -$a_{2}$ 
& 0 & 0 & -$a_{1}$ & 0  
& -$a_{2}$ & 0 & 0 & -$a_{8}$\\
$s_{31}u_{23}^2u_{32}$ 
& 0 & 0 & 0 & -$a_{1}$ 
& 0 & 0 & 0 & 0 
& -$a_{1}$ & $a_{2}$ & 0 & $a_{2}$ \\
$s_{31}u_{23}^2u_{31}$ 
& 0 & 0 & 0 & -$a_{1}$ 
& 0 & 0 & 0 & 0 
& 0 & 0 & 0 & $a_{2}$ \\
$s_{31}u_{23}^3$ 
& $a_{2}$ & 0 & 0 & 0 
& 0 & 0 & -$a_{2}$ & 0 
& $a_{2}$ & 0 & 0 & -$a_{8}$ \\
$s_{31}u_{21}u_{32}^2$ 
& $a_{1}$ & 0 & 0 & $a_{2}$ 
& 0 & 0 & $a_{2}$ & 0 
& 0 & 0 & 0 & 0 \\
$s_{31}u_{21}u_{31}u_{32}$ 
& $a_{2}$ & 0 & 0 & $a_{2}$ 
& 0 & 0 & $a_{2}$ & 0 
& 0 & 0 & 0 & -$a_{2}$ \\
$s_{31}u_{21}u_{31}^2$ 
& $a_{2}$ & 0 & 0 & 0 
& 0 & 0 & 0 & 0 
& $a_{2}$ & 0 & 0 & 0 \\
$s_{31}u_{21}u_{23}u_{32}$ 
& -$a_{2}$ & 0 & 0 & -$a_{2}$ 
& 0 & 0 & -$a_{2}$ & 0 
& -$a_{1}$ & $a_{10}$ & 0 & $a_{2}$ \\
$s_{31}u_{21}u_{23}u_{31}$ 
& $a_{1}$ & 0 & 0 & -$a_{4}$ 
& 0 & 0 & -$a_{4}$ & 0 
& 0 & $a_{8}$ & 0 & $a_{2}$ \\
$s_{31}u_{21}u_{23}^2$ 
& $a_{2}$ & 0 & 0 & -$a_{1}$ 
& 0 & 0 & -$a_{1}$ & 0 
& $a_{2}$ & -$a_{2}$ & 0 & -$a_{2}$ \\
$s_{31}u_{21}^2u_{32}$ 
& -$a_{2}$ & 0 & 0 & 0 
& 0 & 0 & 0 & 0 
& 0 & $a_{8}$ & 0 & 0 \\
$s_{31}u_{21}^2u_{31}$ 
& $a_{1}$ & 0 & 0 & -$a_{2}$ 
& 0 & 0 & 0 & 0 
& $a_{1}$ & 0 & 0 & $a_{2}$ \\
$s_{31}u_{21}^2u_{23}$ 
& $a_{2}$ & 0 & 0 & -$a_{1}$ 
& 0 & 0 & -$a_{1}$ & 0 
& $a_{2}$ & -$a_{2}$ & 0 & -$a_{2}$ \\
$s_{31}u_{21}^3$ 
& $a_{2}$ & 0 & 0 & -$a_{2}$ 
& 0 & 0 & 0 & 0 
& $a_{2}$ & -$a_{8}$ & 0 & 0 \\
$s_{31}u_{13}u_{23}u_{32}$ 
& $a_{1}$ & 0 & 0 & 0 
& 0 & 0 & $a_{2}$ & 0 
& -$a_{1}$ & -$a_{8}$ & 0 & -$a_{1}$ \\
$s_{31}u_{13}u_{23}u_{31}$ 
& 0 & -$a_{7}$ & -$a_{7}$ & 0 
& 0 & 0 & $a_{2}$ & 0 
& -$a_{5}$ & $a_{14}$ & $a_{7}$ & $a_{14}$ \\
$s_{31}u_{13}u_{23}^2$ 
& $a_{1}$ & 0 & 0 & 0 
& 0 & 0 & -$a_{2}$ & 0 
& $a_{1}$ & $a_{2}$ & 0 & $a_{10}$ \\
$s_{31}u_{13}u_{21}u_{32}$ 
& $a_{2}$ & 0 & 0 & 0 
& 0 & 0 & 0 & 0 
& -$a_{5}$ & $a_{10}$ & 0 & -$a_{1}$ \\
$s_{31}u_{13}u_{21}u_{31}$ 
& 0 & -$a_{7}$ & -$a_{7}$ & $a_{2}$ 
& 0 & 0 & 0 & 0 
& -$a_{5}$ & $a_{4}$ & $a_{7}$ & $a_{10}$ \\
$s_{31}u_{13}u_{21}u_{23}$ 
& $a_{1}$ & -$a_{7}$ & -$a_{7}$ & -$a_{4}$ 
& 0 & 0 & -$a_{4}$ & 0 
& 0 & $a_{1}$ & $a_{7}$ & $a_{15}$ \\
$s_{31}u_{13}u_{21}^2$ 
& 0 & -$a_{7}$ & -$a_{7}$ & 0 
& 0 & 0 & -$a_{1}$ & 0 
& 0 & $a_{1}$ & $a_{7}$ & $a_{1}$ \\
$s_{31}u_{13}^2u_{23}$ 
& $a_{2}$ & $a_{7}$ & $a_{7}$ & 0 
& 0 & 0 & 0 & 0 
& $a_{2}$ & -$a_{2}$ & -$a_{7}$ & -$a_{2}$ \\
$s_{31}u_{13}^2u_{21}$ 
& $a_{2}$ & $a_{7}$ & $a_{7}$ & -$a_{1}$ 
& 0 & 0 & -$a_{2}$ & 0 
& -$a_{2}$ & -$a_{10}$ & -$a_{7}$ & -$a_{8}$ \\
$s_{31}u_{12}u_{23}u_{32}$ 
& $a_{3}$ & 0 & 0 & $a_{4}$ 
& 0 & 0 & $a_{1}$ & 0 
& 0 & $a_{1}$ & 0 & $a_{2}$ \\
$s_{31}u_{12}u_{23}u_{31}$ 
& $a_{2}$ & -$a_{7}$ & -$a_{7}$ & 0 
& 0 & 0 & 0 & 0 
& -$a_{5}$ & $a_{2}$ & $a_{7}$ & $a_{16}$ \\
$s_{31}u_{12}u_{23}^2$ 
& -$a_{2}$ & 0 & 0 & 0 
& 0 & 0 & 0 & 0 
& 0 & -$a_{1}$ & 0 & $a_{8}$ \\
$s_{31}u_{12}u_{21}u_{32}$ 
& $a_{3}$ & 0 & 0 & $a_{1}$ 
& 0 & 0 & $a_{4}$ & 0 
& 0 & $a_{2}$ & 0 & $a_{4}$ \\
$s_{31}u_{12}u_{21}u_{31}$ 
& $a_{1}$ & 0 & 0 & $a_{2}$ 
& 0 & 0 & 0 & 0 
& -$a_{1}$ & $a_{2}$ & 0 & $a_{4}$ \\
$s_{31}u_{12}u_{21}u_{23}$ 
& -$a_{2}$ & -$a_{7}$ & -$a_{7}$ & -$a_{2}$ 
& 0 & 0 & -$a_{2}$ & 0 
& -$a_{1}$ & 0 & $a_{7}$ & $a_{14}$ \\
$s_{31}u_{12}u_{21}^2$ 
& 0 & 0 & 0 & 0 
& 0 & 0 & -$a_{1}$ & 0 
& -$a_{1}$ & $a_{2}$ & 0 & $a_{1}$ \\
$s_{31}u_{12}u_{13}u_{23}$ 
& $a_{2}$ & $a_{8}$ & $a_{8}$ & $a_{2}$ 
& 0 & 0 & $a_{2}$ & 0 
& 0 & -$a_{5}$ & -$a_{8}$ & -$a_{16}$ \\
$s_{31}u_{12}u_{13}u_{21}$ 
& 0 & $a_{7}$ & $a_{7}$ & -$a_{1}$ 
& 0 & 0 & 0 & 0 
& -$a_{1}$ & -$a_{4}$ & -$a_{7}$ & -$a_{15}$ \\
$s_{31}u_{12}^2u_{23}$ 
& $a_{1}$ & $a_{7}$ & $a_{7}$ & $a_{2}$ 
& 0 & 0 & $a_{2}$ & 0 
& 0 & 0 & -$a_{7}$ & -$a_{2}$ \\
$s_{31}u_{12}^2u_{21}$ 
& $a_{2}$ & 0 & 0 & -$a_{2}$ 
& 0 & 0 & $a_{2}$ & 0 
& 0 & 0 & 0 & -$a_{2}$ \\
$s_{31}^2u_{23}u_{32}$ 
& $a_{1}$ & $a_{8}$ & -$a_{8}$ & 0 
& $a_{8}$ & $a_{8}$ & 0 & -$a_{8}$ 
& 0 & -$a_{2}$ & $a_{7}$ & 0 \\
$s_{31}^2u_{23}u_{31}$ 
& 0 & $a_{7}$ & -$a_{9}$ & -$a_{1}$ 
& 0 & $a_{2}$ & 0 & 0 
& -$a_{5}$ & $a_{10}$ & -$a_{7}$ & $a_{4}$ \\
$s_{31}^2u_{23}^2$ 
& 0 & -$a_{8}$ & $a_{8}$ & $a_{1}$ 
& -$a_{8}$ & -$a_{8}$ & 0 & $a_{8}$  
& 0 & $a_{1}$ & -$a_{7}$ & $a_{2}$ \\
$s_{31}^2u_{21}u_{32}$ 
& $a_{5}$ & -$a_{7}$ & -$a_{8}$ & 0 
& $a_{2}$ & 0 & 0 & 0 
& 0 & 0 & $a_{7}$ & 0 \\

\hline\hline
\end{tabular}
\end{center}
\end{table}

\begin{table}
\centerline{Table 1. Continued.}
\begin{center}
\begin{tabular}{lllllllllllll}
\hline\hline
& $A$ & $B$ & $C$ & $D$  & $E$  & $F$ 
& $G$ & $H$ & $I$ & $J$ & $K$ & $L$ \\
\hline
$s_{31}^2u_{21}u_{31}$ 
& $a_{1}$ & 0 & -$a_{2}$ & -$a_{2}$ 
& $a_{8}$ & $a_{8}$ & $a_{2}$ & -$a_{8}$ 
& 0 & $a_{8}$ & 0 & $a_{4}$ \\
$s_{31}^2u_{21}u_{23}$ 
& -$a_{1}$ & 0 & $a_{7}$ & $a_{2}$ 
& -$a_{10}$ & $a_{8}$ & $a_{2}$ & $a_{8}$ 
& -$a_{5}$ & $a_{5}$ & -$a_{9}$ & $a_{3}$ \\
$s_{31}^2u_{21}^2$ 
& 0 & $a_{7}$ & -$a_{8}$ & 0 
& -$a_{8}$ & $a_{8}$ & $a_{1}$ & -$a_{8}$ 
& 0 & $a_{2}$ & -$a_{7}$ & $a_{5}$ \\
$s_{31}^2u_{13}u_{23}$ 
& $a_{1}$ & -$a_{8}$ & $a_{7}$ & $a_{2}$ 
& $a_{8}$ & -$a_{8}$ & -$a_{2}$ & $a_{8}$ 
& 0 & 0 & -$a_{7}$ & -$a_{1}$ \\
$s_{31}^2u_{13}u_{21}$ 
& 0 & -$a_{7}$ & $a_{8}$ & 0 
& 0 & 0 & -$a_{1}$ & $a_{2}$ 
& -$a_{5}$ & $a_{2}$ & -$a_{8}$ & -$a_{1}$ \\
$s_{31}^2u_{12}u_{23}$ 
& $a_{5}$ & 0 & -$a_{7}$ & 0 
& $a_{2}$ & 0 & 0 & 0 
& 0 & 0 & 0 & -$a_{2}$ \\
$s_{31}^2u_{12}u_{21}$ 
& $a_{1}$ & -$a_{7}$ & $a_{8}$ & 0 
& $a_{8}$ & -$a_{8}$ & 0 & $a_{8}$ 
& 0 & 0 & $a_{7}$ & -$a_{1}$ \\
$s_{31}^3u_{23}$ 
& $a_{1}$ & -$a_{7}$ & -$a_{8}$ & -$a_{2}$ 
& $a_{2}$ & 0 & -$a_{2}$ & 0 
& 0 & 0 & $a_{7}$ & 0 \\
$s_{31}^3u_{21}$ 
& $a_{1}$ & -$a_{7}$ & -$a_{8}$ & -$a_{2}$ 
& $a_{2}$ & 0 & -$a_{2}$ & 0 
& 0 & 0 & $a_{7}$ & 0 \\
$s_{23}u_{13}u_{32}^2$ 
& $a_{2}$ & -$a_{7}$ & -$a_{7}$ & -$a_{2}$ 
& 0 & 0 & -$a_{2}$ & 0 
& -$a_{1}$ & -$a_{8}$ & $a_{19}$ & 0 \\
$s_{23}u_{13}u_{31}u_{32}$ 
& 0 & -$a_{7}$ & -$a_{7}$ & 0 
& 0 & 0 & -$a_{1}$ & 0 
& -$a_{1}$ & -$a_{2}$ & 0 & -$a_{8}$ \\
$s_{23}u_{13}u_{31}^2$ 
& $a_{2}$ & 0 & 0 & $a_{2}$ 
& 0 & 0 & 0 & 0 
& -$a_{2}$ & 0 & 0 & -$a_{8}$ \\
$s_{23}u_{13}u_{23}u_{32}$ 
& 0 & $a_{8}$ & $a_{8}$ & 0 
& 0 & 0 & -$a_{5}$ & 0 
& $a_{2}$ & $a_{8}$ & -$a_{7}$ & $a_{8}$ \\
$s_{23}u_{13}u_{23}u_{31}$ 
& $a_{1}$ & 0 & 0 & 0 
& 0 & 0 & -$a_{1}$ & 0 
& $a_{2}$ & 0 & $a_{19}$ & 0 \\
$s_{23}u_{13}u_{23}^2$ 
& $a_{2}$ & -$a_{7}$ & -$a_{7}$ & 0
& 0 & 0 & $a_{2}$ & 0 
& 0 & $a_{2}$ & $a_{19}$ & $a_{8}$ \\
$s_{23}u_{13}u_{21}u_{32}$ 
& $a_{2}$ & $a_{7}$ & $a_{7}$ & 0 
& 0 & 0 & -$a_{5}$ & 0 
& 0 & $a_{2}$ & -$a_{7}$ & 0 \\
$s_{23}u_{13}u_{21}u_{31}$ 
& $a_{3}$ & 0 & 0 & $a_{4}$ 
& 0 & 0 & 0 & 0 
& $a_{1}$ & $a_{1}$ & $a_{19}$ & 0 \\
$s_{23}u_{13}u_{21}u_{23}$ 
& $a_{2}$ & -$a_{8}$ & -$a_{8}$ & $a_{2}$ 
& 0 & 0 & 0 & 0 
& $a_{2}$ & $a_{2}$ & $a_{22}$ & 0 \\
$s_{23}u_{13}u_{21}^2$ 
& $a_{1}$ & -$a_{7}$ & -$a_{7}$ & $a_{2}$ 
& 0 & 0 & 0 & 0 
& $a_{2}$ & $a_{2}$ & $a_{7}$ & 0 \\
$s_{23}u_{13}^2u_{32}$ 
& 0 & $a_{7}$ & $a_{7}$ & -$a_{1}$ 
& 0 & 0 & 0 & 0 
& 0 & $a_{10}$ & 0 & -$a_{10}$ \\
$s_{23}u_{13}^2u_{31}$ 
& 0 & 0 & 0 & -$a_{1}$ 
& 0 & 0 & -$a_{1}$ & 0 
& 0 & $a_{1}$ & 0 & $a_{2}$ \\
$s_{23}u_{13}^2u_{23}$ 
& $a_{1}$ & 0 & 0 & 0 
& 0 & 0 & $a_{1}$ & 0
& -$a_{2}$ & $a_{2}$ & -$a_{19}$ & $a_{10}$ \\
$s_{23}u_{13}^2u_{21}$ 
& -$a_{2}$ & 0 & 0 & 0 
& 0 & 0 & 0 & 0 
& 0 & $a_{2}$ & -$a_{19}$ & $a_{2}$ \\
$s_{23}u_{13}^3$ 
& $a_{2}$ & 0 & 0 & 0 
& 0 & 0 & $a_{2}$ & 0 
& -$a_{2}$ & -$a_{2}$ & 0 & 0 \\
$s_{23}u_{12}u_{32}^2$ 
& $a_{2}$ & -$a_{7}$ & -$a_{7}$ & 0 
& 0 & 0 & $a_{2}$ & 0 
& 0 & $a_{8}$ & $a_{19}$ & $a_{2}$ \\
$s_{23}u_{12}u_{31}u_{32}$ 
& $a_{2}$ & -$a_{8}$ & -$a_{8}$ & $a_{2}$ 
& 0 & 0 & 0 & 0 
& $a_{2}$ & 0 & $a_{22}$ & $a_{2}$ \\
$s_{23}u_{12}u_{31}^2$ 
& $a_{1}$ & -$a_{7}$ & -$a_{7}$ & $a_{2}$ 
& 0 & 0 & 0 & 0 
& $a_{2}$ & 0 & $a_{7}$ & $a_{2}$ \\
$s_{23}u_{12}u_{23}u_{32}$ 
& 0 & $a_{8}$ & $a_{8}$ & $a_{2}$ 
& 0 & 0 & -$a_{5}$ & 0 
& 0 & $a_{8}$ & -$a_{7}$ & $a_{8}$ \\
$s_{23}u_{12}u_{23}u_{31}$ 
& $a_{2}$ & $a_{7}$ & $a_{7}$ & 0 
& 0 & 0 & -$a_{5}$ & 0 
& 0 & 0 & -$a_{7}$ & $a_{2}$ \\
$s_{23}u_{12}u_{23}^2$ 
& $a_{2}$ & -$a_{7}$ & -$a_{7}$ & -$a_{1}$ 
& 0 & 0 & -$a_{2}$ & 0 
& -$a_{2}$ & 0 & $a_{19}$ & -$a_{8}$ \\
$s_{23}u_{12}u_{21}u_{32}$ 
& $a_{1}$ & 0 & 0 & $a_{2}$ 
& 0 & 0 & -$a_{1}$ & 0 
& 0 & 0 & $a_{19}$ & 0 \\
$s_{23}u_{12}u_{21}u_{31}$ 
& $a_{3}$ & 0 & 0 & $a_{1}$ 
& 0 & 0 & 0 & 0 
& $a_{4}$ & 0 & $a_{19}$ & $a_{1}$ \\
$s_{23}u_{12}u_{21}u_{23}$ 
& 0 & -$a_{7}$ & -$a_{7}$ & -$a_{1}$ 
& 0 & 0 & -$a_{1}$ & 0 
& 0 & -$a_{8}$ & 0 & -$a_{2}$ \\
$s_{23}u_{12}u_{21}^2$ 
& $a_{2}$ & 0 & 0 & -$a_{2}$ 
& 0 & 0 & 0 & 0 
& $a_{2}$ & -$a_{8}$ & 0 & 0 \\
$s_{23}u_{12}u_{13}u_{32}$ 
& $a_{1}$ & $a_{7}$ & $a_{7}$ & -$a_{4}$ 
& 0 & 0 & 0 & 0 
& -$a_{4}$ & $a_{2}$ & -$a_{19}$ & -$a_{8}$ \\
$s_{23}u_{12}u_{13}u_{31}$ 
& -$a_{2}$ & $a_{7}$ & $a_{7}$ & -$a_{2}$ 
& 0 & 0 & -$a_{1}$ & 0 
& -$a_{2}$ & $a_{1}$ & -$a_{8}$ & $a_{2}$ \\
$s_{23}u_{12}u_{13}u_{23}$ 
& $a_{1}$ & $a_{7}$ & $a_{7}$ & -$a_{4}$ 
& 0 & 0 & 0 & 0 
& -$a_{4}$ & -$a_{8}$ & -$a_{19}$ & $a_{2}$ \\
$s_{23}u_{12}u_{13}u_{21}$ 
& -$a_{2}$ & $a_{7}$ & $a_{7}$ & -$a_{2}$ 
& 0 & 0 & -$a_{1}$ & 0 
& -$a_{2}$ & $a_{2}$ & -$a_{8}$ & $a_{1}$ \\
$s_{23}u_{12}u_{13}^2$ 
& $a_{2}$ & 0 & 0 & -$a_{1}$ 
& 0 & 0 & $a_{2}$ & 0 
& -$a_{1}$ & -$a_{10}$ & $a_{7}$ & -$a_{10}$ \\

\hline\hline
\end{tabular}
\end{center}
\end{table}

\begin{table}
\centerline{Table 1. Continued.}
\begin{center}
\begin{tabular}{lllllllllllll}
\hline\hline
& $A$ & $B$ & $C$ & $D$  & $E$  & $F$ 
& $G$ & $H$ & $I$ & $J$ & $K$ & $L$ \\
\hline
$s_{23}u_{12}^2u_{32}$ 
& $a_{1}$ & 0 & 0 & -$a_{2}$ 
& 0 & 0 & $a_{1}$ & 0 
& 0 & $a_{10}$ & -$a_{19}$ & $a_{2}$ \\
$s_{23}u_{12}^2u_{31}$ 
& -$a_{2}$ & 0 & 0 & 0 
& 0 & 0 & 0 & 0 
& 0 & $a_{2}$ & -$a_{19}$ & $a_{2}$ \\
$s_{23}u_{12}^2u_{23}$ 
& 0 & $a_{7}$ & $a_{7}$ & 0 
& 0 & 0 & 0 & 0 
& -$a_{1}$ & -$a_{10}$ & 0 & $a_{10}$ \\
$s_{23}u_{12}^2u_{21}$ 
& 0 & 0 & 0 & 0 
& 0 & 0 & -$a_{1}$ & 0 
& -$a_{1}$ & $a_{2}$ & 0 & $a_{1}$ \\
$s_{23}u_{12}^2u_{13}$ 
& $a_{2}$ & 0 & 0 & -$a_{1}$ 
& 0 & 0 & $a_{2}$ & 0 
& -$a_{1}$ & -$a_{10}$ & $a_{7}$ & -$a_{10}$ \\
$s_{23}u_{12}^3$ 
& $a_{2}$ & 0 & 0 & -$a_{2}$ 
& 0 & 0 & $a_{2}$ & 0 
& 0 & 0 & 0 & -$a_{2}$ \\
$s_{23}s_{31}u_{32}^2$ 
& 0 & $a_{8}$ & 0 & -$a_{2}$ 
& $a_{2}$ & 0 & -$a_{2}$ & 0 
& 0 & -$a_{8}$ & 0 & 0 \\
$s_{23}s_{31}u_{31}u_{32}$ 
& -$a_{1}$ & $a_{7}$ & $a_{9}$ & -$a_{1}$ 
& $a_{2}$ & $a_{2}$ & 0 & -$a_{2}$ 
& 0 & 0 & -$a_{22}$ & -$a_{2}$ \\
$s_{23}s_{31}u_{31}^2$ 
& 0 & -$a_{7}$ & $a_{9}$ & -$a_{2}$ 
& 0 & $a_{2}$ & 0 & 0 
& -$a_{2}$ & 0 & $a_{7}$ & -$a_{2}$ \\
$s_{23}s_{31}u_{23}u_{32}$ 
& $a_{1}$ & -$a_{9}$ & -$a_{7}$ & $a_{1}$ 
& -$a_{10}$ & $a_{8}$ & 0 & -$a_{8}$ 
& -$a_{1}$ & $a_{10}$ & $a_{19}$ & $a_{2}$ \\
$s_{23}s_{31}u_{23}u_{31}$ 
& $a_{5}$ & -$a_{7}$ & -$a_{11}$ & 0 
& -$a_{2}$ & 0 & -$a_{5}$ & $a_{2}$ 
& -$a_{1}$ & 0 & $a_{7}$ & $a_{2}$ \\
$s_{23}s_{31}u_{23}^2$ 
& 0 & $a_{7}$ & $a_{7}$ & -$a_{1}$ 
& $a_{8}$ & -$a_{8}$ & -$a_{4}$ & $a_{8}$ 
& $a_{2}$ & 0 & -$a_{19}$ & 0 \\
$s_{23}s_{31}u_{21}u_{32}$ 
& $a_{2}$ & -$a_{2}$ & $a_{8}$ & 0 
& 0 & $a_{2}$ & $a_{1}$ & -$a_{2}$ 
& 0 & $a_{1}$ & -$a_{7}$ & 0 \\
$s_{23}s_{31}u_{21}u_{31}$ 
& $a_{3}$ & -$a_{9}$ & 0 & -$a_{2}$ 
& -$a_{8}$ & $a_{10}$ & 0 & $a_{8}$ 
& 0 & 0 & $a_{20}$ & 0 \\
$s_{23}s_{31}u_{21}u_{23}$ 
& $a_{2}$ & $a_{9}$ & -$a_{9}$ & -$a_{2}$ 
& -$a_{8}$ & -$a_{8}$ & -$a_{6}$ & $a_{10}$ 
& 0 & -$a_{10}$ & 0 & 0 \\
$s_{23}s_{31}u_{21}^2$ 
& $a_{4}$ & $a_{7}$ & -$a_{8}$ & -$a_{2}$ 
& -$a_{8}$ & $a_{8}$ & 0 & $a_{8}$ 
& $a_{2}$ & -$a_{10}$ & $a_{8}$ & 0 \\
$s_{23}s_{31}u_{13}u_{32}$ 
& $a_{5}$ & -$a_{2}$ & -$a_{8}$ & 0 
& 0 & -$a_{2}$ & -$a_{1}$ & $a_{2}$ 
& -$a_{5}$ & $a_{2}$ & $a_{7}$ & -$a_{1}$ \\
$s_{23}s_{31}u_{13}u_{31}$ 
& $a_{1}$ & -$a_{8}$ & -$a_{10}$ & $a_{1}$ 
& $a_{8}$ & -$a_{10}$ & -$a_{1}$ & -$a_{8}$ 
& 0 & $a_{1}$ & $a_{7}$ & $a_{16}$ \\
$s_{23}s_{31}u_{13}u_{23}$ 
& $a_{1}$ & $a_{7}$ & 0 & -$a_{6}$ 
& $a_{2}$ & $a_{2}$ & -$a_{1}$ & 0 
& -$a_{1}$ & $a_{15}$ & -$a_{21}$ & $a_{17}$ \\
$s_{23}s_{31}u_{13}u_{21}$ 
& $a_{2}$ & -$a_{7}$ & -$a_{11}$ & -$a_{1}$ 
& $a_{2}$ & 0 & -$a_{6}$ & 0 
& 0 & $a_{3}$ & -$a_{8}$ & $a_{18}$ \\
$s_{23}s_{31}u_{13}^2$ 
& 0 & $a_{9}$ & $a_{9}$ & -$a_{1}$ 
& -$a_{8}$ & $a_{8}$ & $a_{2}$ & $a_{8}$ 
& -$a_{4}$ & -$a_{8}$ & -$a_{8}$ & -$a_{10}$ \\
$s_{23}s_{31}u_{12}u_{32}$ 
& $a_{3}$ & -$a_{7}$ & -$a_{11}$ & -$a_{2}$ 
& $a_{10}$ & -$a_{8}$ & 0 & $a_{8}$ 
& 0 & $a_{2}$ & $a_{21}$ & $a_{4}$ \\
$s_{23}s_{31}u_{12}u_{31}$ 
& $a_{2}$ & 0 & -$a_{12}$ & 0 
& $a_{2}$ & 0 & 0 & -$a_{2}$ 
& $a_{1}$ & $a_{1}$ & -$a_{7}$ & $a_{6}$ \\
$s_{23}s_{31}u_{12}u_{23}$ 
& $a_{2}$ & -$a_{7}$ & $a_{8}$ & -$a_{1}$ 
& 0 & $a_{2}$ & 0 & 0 
& -$a_{6}$ & $a_{4}$ & -$a_{20}$ & $a_{15}$ \\
$s_{23}s_{31}u_{12}u_{21}$ 
& 0 & 0 & 0 & 0 
& 0 & 0 & 0 & -$a_{2}$ 
& 0 & $a_{5}$ & -$a_{9}$ & $a_{18}$ \\
$s_{23}s_{31}u_{12}u_{13}$ 
& $a_{2}$ & $a_{7}$ & $a_{1}$ & -$a_{2}$ 
& -$a_{8}$ & -$a_{8}$ & 0 & $a_{10}$ 
& -$a_{6}$ & -$a_{2}$ & -$a_{21}$ & -$a_{17}$ \\
$s_{23}s_{31}u_{12}^2$ 
& $a_{4}$ & -$a_{7}$ & $a_{8}$ & -$a_{2}$ 
& $a_{8}$ & -$a_{8}$ & $a_{2}$ & $a_{8}$ 
& 0 & 0 & $a_{7}$ & -$a_{4}$ \\
$s_{23}s_{31}^2u_{32}$ 
& $a_{1}$ & $a_{7}$ & -$a_{8}$ & -$a_{1}$ 
& $a_{1}$ & 0 & -$a_{1}$ & 0 
& 0 & -$a_{8}$ & $a_{7}$ & 0 \\
$s_{23}s_{31}^2u_{31}$ 
& 0 & 0 & -$a_{9}$ & -$a_{1}$ 
& $a_{2}$ & $a_{2}$ & 0 & -$a_{2}$ 
& 0 & $a_{2}$ & -$a_{19}$ & $a_{4}$ \\
$s_{23}s_{31}^2u_{23}$ 
& $a_{1}$ & -$a_{2}$ & -$a_{8}$ & 0 
& -$a_{8}$ & $a_{8}$ & 0 & $a_{8}$ 
& -$a_{5}$ & $a_{15}$ & -$a_{22}$ & $a_{3}$ \\
$s_{23}s_{31}^2u_{21}$ 
& $a_{1}$ & -$a_{8}$ & -$a_{10}$ & -$a_{1}$ 
& $a_{8}$ & $a_{10}$ & $a_{1}$ & -$a_{10}$ 
& 0 & $a_{4}$ & -$a_{7}$ & $a_{18}$ \\
$s_{23}s_{31}^2u_{13}$ 
& $a_{1}$ & -$a_{7}$ & $a_{7}$ & $a_{2}$ 
& -$a_{8}$ & -$a_{8}$ & -$a_{2}$ & $a_{10}$ 
& -$a_{5}$ & $a_{8}$ & -$a_{22}$ & -$a_{1}$ \\
$s_{23}s_{31}^2u_{12}$ 
& $a_{5}$ & -$a_{8}$ & 0 & -$a_{2}$ 
& $a_{10}$ & -$a_{8}$ & -$a_{2}$ & $a_{8}$ 
& 0 & 0 & $a_{8}$ & -$a_{1}$ \\
$s_{23}s_{31}^3$ 
& $a_{1}$ & -$a_{7}$ & -$a_{8}$ & -$a_{2}$ 
& $a_{2}$ & 0 & -$a_{2}$ & 0 
& 0 & 0 & $a_{7}$ & 0 \\
$s_{23}^2u_{13}u_{32}$ 
& 0 & -$a_{7}$ & $a_{8}$ & -$a_{1}$ 
& $a_{2}$ & 0 & -$a_{5}$ & 0 
& 0 & 0 & -$a_{22}$ & 0 \\
$s_{23}^2u_{13}u_{31}$ 
& $a_{1}$ & -$a_{8}$ & $a_{7}$ & 0 
& $a_{8}$ & $a_{8}$ & 0 & -$a_{8}$ 
& 0 & 0 & $a_{8}$ & -$a_{2}$ \\
$s_{23}^2u_{13}u_{23}$ 
& $a_{1}$ & -$a_{7}$ & -$a_{2}$ & $a_{2}$ 
& -$a_{8}$ & $a_{8}$ & 0 & $a_{8}$ 
& -$a_{2}$ & $a_{2}$ & $a_{7}$ & $a_{2}$ \\
$s_{23}^2u_{13}u_{21}$ 
& $a_{5}$ & -$a_{9}$ & -$a_{9}$ & 0 
& 0 & $a_{2}$ & 0 & 0 
& 0 & $a_{2}$ & $a_{9}$ & 0 \\
$s_{23}^2u_{13}^2$ 
& 0 & $a_{8}$ & -$a_{7}$ & $a_{1}$ 
& -$a_{8}$ & -$a_{8}$ & 0 & $a_{8}$ 
& 0 & $a_{1}$ & -$a_{8}$ & $a_{1}$ \\

\hline\hline
\end{tabular}
\end{center}
\end{table}

\begin{table}
\centerline{Table 1. Continued.}
\begin{center}
\begin{tabular}{lllllllllllll}
\hline\hline
& $A$ & $B$ & $C$ & $D$  & $E$  & $F$ 
& $G$ & $H$ & $I$ & $J$ & $K$ & $L$ \\
\hline
$s_{23}^2u_{12}u_{32}$ 
& $a_{1}$ & -$a_{2}$ & -$a_{7}$ & -$a_{2}$ 
& $a_{8}$ & $a_{8}$ & 0 & -$a_{8}$ 
& $a_{2}$ & $a_{2}$ & $a_{7}$ & $a_{2}$ \\
$s_{23}^2u_{12}u_{31}$ 
& $a_{5}$ & -$a_{9}$ & -$a_{9}$ & 0 
& 0 & $a_{2}$ & 0 & 0 
& 0 & 0 & $a_{9}$ & $a_{2}$ \\
$s_{23}^2u_{12}u_{23}$ 
& 0 & $a_{8}$ & -$a_{7}$ & 0 
& 0 & 0 & -$a_{5}$ & $a_{2}$ 
& -$a_{1}$ & 0 & -$a_{22}$ & 0 \\
$s_{23}^2u_{12}u_{21}$ 
& $a_{1}$ & $a_{7}$ & -$a_{8}$ & 0 
& -$a_{8}$ & $a_{8}$ & 0 & $a_{8}$ 
& 0 & -$a_{2}$ & $a_{8}$ & 0 \\
$s_{23}^2u_{12}u_{13}$ 
& -$a_{1}$ & $a_{2}$ & $a_{2}$ & $a_{2}$ 
& $a_{8}$ & -$a_{10}$ & -$a_{5}$ & $a_{8}$ 
& $a_{2}$ & $a_{4}$ & -$a_{23}$ & $a_{4}$ \\
$s_{23}^2u_{12}^2$ 
& 0 & -$a_{7}$ & $a_{8}$ & 0 
& $a_{8}$ & -$a_{8}$ & 0 & -$a_{8}$ 
& $a_{1}$ & $a_{1}$ & -$a_{8}$ & $a_{1}$ \\
$s_{23}^2s_{31}u_{32}$ 
& 0 & -$a_{8}$ & $a_{8}$ & -$a_{1}$ 
& $a_{2}$ & $a_{2}$ & 0 & -$a_{2}$ 
& 0 & $a_{2}$ & -$a_{7}$ & 0 \\
$s_{23}^2s_{31}u_{31}$ 
& $a_{1}$ & -$a_{9}$ & $a_{7}$ & -$a_{1}$ 
& 0 & $a_{1}$ & 0 & 0 
& -$a_{1}$ & 0 & $a_{9}$ & -$a_{2}$ \\
$s_{23}^2s_{31}u_{23}$ 
& $a_{1}$ & 0 & -$a_{2}$ & $a_{2}$ 
& -$a_{8}$ & -$a_{8}$ & -$a_{5}$ & $a_{10}$ 
& -$a_{2}$ & 0 & $a_{19}$ & $a_{2}$ \\
$s_{23}^2s_{31}u_{21}$ 
& $a_{5}$ & -$a_{7}$ & -$a_{2}$ & -$a_{2}$ 
& -$a_{8}$ & $a_{10}$ & 0 & $a_{8}$ 
& -$a_{2}$ & -$a_{2}$ & $a_{2}$ & 0 \\
$s_{23}^2s_{31}u_{13}$ 
& $a_{1}$ & 0 & -$a_{11}$ & 0 
& $a_{8}$ & -$a_{8}$ & -$a_{5}$ & $a_{8}$ 
& 0 & $a_{4}$ & -$a_{21}$ & $a_{6}$ \\
$s_{23}^2s_{31}u_{12}$ 
& $a_{1}$ & -$a_{2}$ & -$a_{8}$ & -$a_{1}$ 
& $a_{10}$ & $a_{8}$ & 0 & -$a_{10}$ 
& $a_{1}$ & $a_{5}$ & -$a_{8}$ & $a_{6}$ \\
$s_{23}^2s_{31}^2$ 
& $a_{1}$ & -$a_{9}$ & -$a_{2}$ & -$a_{1}$ 
& $a_{2}$ & $a_{2}$ & 0 & -$a_{2}$ 
& 0 & $a_{1}$ & 0 & $a_{5}$ \\
$s_{23}^3u_{13}$ 
& $a_{1}$ & -$a_{8}$ & -$a_{8}$ & -$a_{2}$ 
& 0 & $a_{2}$ & 0 & 0 
& -$a_{2}$ & 0 & $a_{8}$ & 0 \\
$s_{23}^3u_{12}$ 
& $a_{1}$ & -$a_{8}$ & -$a_{8}$ & -$a_{2}$ 
& 0 & $a_{2}$ & 0 & 0 
& -$a_{2}$ & 0 & $a_{8}$ & 0 \\
$s_{23}^3s_{31}$ 
& $a_{1}$ & -$a_{8}$ & -$a_{8}$ & -$a_{2}$ 
& 0 & $a_{2}$ & 0 & 0 
& -$a_{2}$ & 0 & $a_{8}$ & 0 \\
$s_{12}u_{32}^3$ 
& $a_{2}$ & 0 & 0 & $a_{2}$ 
& 0 & 0 & -$a_{2}$ & 0 
& 0 & -$a_{8}$ & 0 & 0 \\
$s_{12}u_{31}u_{32}^2$ 
& $a_{2}$ & 0 & 0 & $a_{2}$ 
& 0 & 0 & -$a_{1}$ & 0 
& -$a_{1}$ & -$a_{2}$ & 0 & -$a_{2}$ \\
$s_{12}u_{31}^2u_{32}$ 
& $a_{2}$ & 0 & 0 & $a_{2}$ 
& 0 & 0 & -$a_{1}$ & 0 
& -$a_{1}$ & -$a_{2}$ & 0 & -$a_{2}$ \\
$s_{12}u_{31}^3$ 
& $a_{2}$ & 0 & 0 & $a_{2}$ 
& 0 & 0 & 0 & 0 
& -$a_{2}$ & 0 & 0 & -$a_{8}$ \\
$s_{12}u_{23}u_{32}^2$ 
& 0 & 0 & 0 & -$a_{1}$ 
& 0 & 0 & 0 & 0 
& -$a_{1}$ & $a_{2}$ & 0 & $a_{2}$ \\
$s_{12}u_{23}u_{31}u_{32}$ 
& -$a_{2}$ & 0 & 0 & -$a_{1}$ 
& 0 & 0 & -$a_{2}$ & 0 
& -$a_{2}$ & $a_{2}$ & 0 & $a_{10}$ \\
$s_{12}u_{23}u_{31}^2$ 
& -$a_{2}$ & 0 & 0 & 0 
& 0 & 0 & 0 & 0 
& 0 & 0 & 0 & $a_{8}$ \\
$s_{12}u_{23}^2u_{32}$ 
& $a_{2}$ & 0 & 0 & 0 
& 0 & 0 & -$a_{2}$ & 0 
& $a_{2}$ & 0 & 0 & -$a_{8}$ \\
$s_{12}u_{23}^2u_{31}$ 
& $a_{1}$ & 0 & 0 & 0 
& 0 & 0 & $a_{2}$ & 0 
& $a_{2}$ & 0 & 0 & 0 \\
$s_{12}u_{21}u_{32}^2$ 
& 0 & 0 & 0 & 0 
& 0 & 0 & 0 & 0 
& -$a_{1}$ & $a_{2}$ & 0 & 0 \\
$s_{12}u_{21}u_{31}u_{32}$ 
& $a_{1}$ & 0 & 0 & 0 
& 0 & 0 & -$a_{4}$ & 0 
& -$a_{4}$ & $a_{2}$ & 0 & $a_{8}$ \\
$s_{12}u_{21}u_{31}^2$ 
& $a_{1}$ & 0 & 0 & $a_{1}$ 
& 0 & 0 & 0 & 0 
& -$a_{2}$ & $a_{2}$ & 0 & 0 \\
$s_{12}u_{21}u_{23}u_{32}$ 
& 0 & 0 & 0 & -$a_{1}$ 
& 0 & 0 & -$a_{1}$ & 0 
& 0 & -$a_{2}$ & 0 & -$a_{8}$ \\
$s_{12}u_{21}u_{23}u_{31}$ 
& $a_{2}$ & 0 & 0 & 0 
& 0 & 0 & $a_{2}$ & 0 
& $a_{2}$ & -$a_{2}$ & 0 & 0 \\
$s_{12}u_{21}^2u_{32}$ 
& $a_{2}$ & 0 & 0 & -$a_{2}$ 
& 0 & 0 & -$a_{1}$ & 0 
& -$a_{2}$ & -$a_{8}$ & 0 & 0 \\
$s_{12}u_{21}^2u_{31}$ 
& $a_{2}$ & 0 & 0 & $a_{2}$ 
& 0 & 0 & 0 & 0 
& 0 & 0 & 0 & 0 \\
$s_{12}u_{13}u_{32}^2$ 
& -$a_{2}$ & 0 & 0 & 0 
& 0 & 0 & 0 & 0 
& 0 & $a_{8}$ & 0 & -$a_{1}$ \\
$s_{12}u_{13}u_{31}u_{32}$ 
& -$a_{2}$ & -$a_{7}$ & -$a_{7}$ & -$a_{1}$ 
& 0 & 0 & -$a_{2}$ & 0 
& -$a_{2}$ & $a_{14}$ & $a_{7}$ & 0 \\
$s_{12}u_{13}u_{31}^2$ 
& 0 & 0 & 0 & -$a_{1}$ 
& 0 & 0 & -$a_{1}$ & 0 
& 0 & $a_{1}$ & 0 & $a_{2}$ \\
$s_{12}u_{13}u_{23}u_{32}$ 
& $a_{3}$ & 0 & 0 & 0 
& 0 & 0 & $a_{1}$ & 0 
& $a_{4}$ & $a_{2}$ & 0 & $a_{1}$ \\
$s_{12}u_{13}u_{23}u_{31}$ 
& $a_{3}$ & 0 & 0 & 0 
& 0 & 0 & $a_{4}$ & 0 
& $a_{1}$ & $a_{4}$ & 0 & $a_{2}$ \\
$s_{12}u_{13}u_{21}u_{32}$ 
& $a_{2}$ & -$a_{7}$ & -$a_{7}$ & -$a_{5}$ 
& 0 & 0 & 0 & 0 
& 0 & $a_{16}$ & $a_{7}$ & $a_{2}$ \\
$s_{12}u_{13}u_{21}u_{31}$ 
& $a_{1}$ & 0 & 0 & -$a_{1}$ 
& 0 & 0 & 0 & 0 
& $a_{2}$ & $a_{4}$ & 0 & $a_{2}$ \\

\hline\hline
\end{tabular}
\end{center}
\end{table}

\begin{table}
\centerline{Table 1. Continued.}
\begin{center}
\begin{tabular}{lllllllllllll}
\hline\hline
& $A$ & $B$ & $C$ & $D$  & $E$  & $F$ 
& $G$ & $H$ & $I$ & $J$ & $K$ & $L$ \\
\hline
$s_{12}u_{13}^2u_{32}$ 
& $a_{1}$ & $a_{7}$ & $a_{7}$ & 0 
& 0 & 0 & $a_{2}$ & 0 
& $a_{2}$ & -$a_{2}$ & -$a_{7}$ & 0 \\
$s_{12}u_{13}^2u_{31}$ 
& $a_{2}$ & 0 & 0 & 0 
& 0 & 0 & $a_{2}$ & 0 
& -$a_{2}$ & -$a_{2}$ & 0 & 0 \\
$s_{12}u_{12}u_{32}^2$ 
& $a_{1}$ & 0 & 0 & $a_{1}$ 
& 0 & 0 & -$a_{2}$ & 0 
& 0 & $a_{10}$ & 0 & $a_{2}$ \\
$s_{12}u_{12}u_{31}u_{32}$ 
& $a_{1}$ & -$a_{7}$ & -$a_{7}$ & 0 
& 0 & 0 & -$a_{4}$ & 0 
& -$a_{4}$ & $a_{15}$ & $a_{7}$ & $a_{1}$ \\
$s_{12}u_{12}u_{31}^2$ 
& 0 & -$a_{7}$ & -$a_{7}$ & 0 
& 0 & 0 & -$a_{1}$ & 0 
& 0 & $a_{1}$ & $a_{7}$ & $a_{1}$ \\
$s_{12}u_{12}u_{23}u_{32}$ 
& $a_{1}$ & 0 & 0 & -$a_{1}$ 
& 0 & 0 & $a_{2}$ & 0 
& 0 & -$a_{1}$ & 0 & -$a_{8}$ \\
$s_{12}u_{12}u_{23}u_{31}$ 
& $a_{2}$ & 0 & 0 & -$a_{5}$ 
& 0 & 0 & 0 & 0 
& 0 & -$a_{1}$ & 0 & $a_{10}$ \\
$s_{12}u_{12}u_{21}u_{32}$ 
& 0 & -$a_{7}$ & -$a_{7}$ & -$a_{5}$ 
& 0 & 0 & $a_{2}$ & 0 
& 0 & $a_{14}$ & $a_{7}$ & $a_{14}$ \\
$s_{12}u_{12}u_{21}u_{31}$ 
& 0 & -$a_{7}$ & -$a_{7}$ & -$a_{5}$ 
& 0 & 0 & 0 & 0 
& $a_{2}$ & $a_{10}$ & $a_{7}$ & $a_{4}$ \\
$s_{12}u_{12}u_{13}u_{32}$ 
& $a_{2}$ & $a_{8}$ & $a_{8}$ & 0 
& 0 & 0 & $a_{2}$ & 0 
& $a_{2}$ & -$a_{16}$ & -$a_{8}$ & -$a_{5}$ \\
$s_{12}u_{12}u_{13}u_{31}$ 
& 0 & $a_{7}$ & $a_{7}$ & -$a_{1}$ 
& 0 & 0 & 0 & 0 
& -$a_{1}$ & -$a_{15}$ & -$a_{7}$ & -$a_{4}$ \\
$s_{12}u_{12}^2u_{32}$ 
& $a_{2}$ & $a_{7}$ & $a_{7}$ & $a_{2}$ 
& 0 & 0 & 0 & 0 
& 0 & -$a_{2}$ & -$a_{7}$ & -$a_{2}$ \\
$s_{12}u_{12}^2u_{31}$ 
& $a_{2}$ & $a_{7}$ & $a_{7}$ & -$a_{2}$ 
& 0 & 0 & -$a_{2}$ & 0 
& -$a_{1}$ & -$a_{8}$ & -$a_{7}$ & -$a_{10}$ \\
$s_{12}s_{31}u_{32}^2$ 
& $a_{4}$ & $a_{8}$ & -$a_{8}$ & $a_{2}$ 
& $a_{8}$ & $a_{8}$ & -$a_{2}$ & -$a_{8}$ 
& 0 & -$a_{10}$ & $a_{7}$ & 0 \\
$s_{12}s_{31}u_{31}u_{32}$ 
& $a_{2}$ & $a_{2}$ & -$a_{9}$ & 0 
& -$a_{8}$ & $a_{10}$ & -$a_{2}$ & -$a_{8}$ 
& -$a_{6}$ & -$a_{8}$ & -$a_{7}$ & $a_{1}$ \\
$s_{12}s_{31}u_{31}^2$ 
& 0 & $a_{7}$ & $a_{7}$ & $a_{2}$ 
& -$a_{8}$ & $a_{8}$ & -$a_{1}$ & $a_{8}$ 
& -$a_{4}$ & $a_{1}$ & -$a_{7}$ & $a_{2}$ \\
$s_{12}s_{31}u_{23}u_{32}$ 
& 0 & 0 & 0 & 0 
& 0 & -$a_{2}$ & 0 & 0 
& 0 & $a_{5}$ & -$a_{8}$ & $a_{5}$ \\
$s_{12}s_{31}u_{23}u_{31}$ 
& $a_{2}$ & -$a_{2}$ & -$a_{7}$ & -$a_{6}$ 
& $a_{2}$ & 0 & -$a_{1}$ & 0 
& 0 & $a_{6}$ & -$a_{8}$ & $a_{15}$ \\
$s_{12}s_{31}u_{23}^2$ 
& $a_{4}$ & -$a_{8}$ & $a_{8}$ & 0 
& -$a_{8}$ & $a_{8}$ & -$a_{2}$ & $a_{8}$ 
& $a_{2}$ & 0 & $a_{7}$ & -$a_{10}$ \\
$s_{12}s_{31}u_{21}u_{32}$ 
& $a_{2}$ & -$a_{7}$ & -$a_{2}$ & 0 
& 0 & 0 & -$a_{1}$ & $a_{2}$ 
& -$a_{6}$ & $a_{15}$ & -$a_{8}$ & $a_{6}$ \\
$s_{12}s_{31}u_{21}u_{31}$ 
& $a_{1}$ & -$a_{7}$ & -$a_{7}$ & -$a_{1}$ 
& $a_{2}$ & 0 & -$a_{6}$ & $a_{2}$ 
& -$a_{1}$ & $a_{17}$ & -$a_{8}$ & $a_{17}$ \\
$s_{12}s_{31}u_{21}u_{23}$ 
& $a_{2}$ & -$a_{9}$ & $a_{2}$ & -$a_{6}$ 
& -$a_{8}$ & $a_{10}$ & -$a_{2}$ & -$a_{8}$ 
& 0 & $a_{1}$ & -$a_{7}$ & -$a_{8}$ \\
$s_{12}s_{31}u_{21}^2$ 
& 0 & $a_{7}$ & $a_{7}$ & -$a_{4}$ 
& $a_{8}$ & $a_{8}$ & -$a_{1}$ & -$a_{8}$ 
& $a_{2}$ & $a_{2}$ & -$a_{7}$ & $a_{1}$ \\
$s_{12}s_{31}u_{13}u_{32}$ 
& $a_{2}$ & -$a_{2}$ & $a_{9}$ & 0 
& $a_{2}$ & -$a_{2}$ & 0 & 0 
& $a_{1}$ & $a_{14}$ & -$a_{9}$ & -$a_{5}$ \\
$s_{12}s_{31}u_{13}u_{31}$ 
& $a_{1}$ & -$a_{10}$ & -$a_{8}$ & -$a_{1}$ 
& $a_{8}$ & -$a_{8}$ & $a_{1}$ & -$a_{10}$ 
& 0 & $a_{14}$ & $a_{8}$ & 0 \\
$s_{12}s_{31}u_{13}u_{23}$ 
& $a_{3}$ & $a_{7}$ & -$a_{8}$ & 0 
& -$a_{8}$ & $a_{8}$ & -$a_{2}$ & $a_{10}$ 
& 0 & -$a_{1}$ & $a_{8}$ & $a_{1}$ \\
$s_{12}s_{31}u_{13}u_{21}$ 
& $a_{5}$ & -$a_{10}$ & -$a_{8}$ & -$a_{5}$ 
& -$a_{2}$ & $a_{2}$ & 0 & 0 
& -$a_{1}$ & $a_{1}$ & $a_{8}$ & $a_{2}$ \\
$s_{12}s_{31}u_{13}^2$ 
& 0 & $a_{11}$ & $a_{7}$ & 0 
& 0 & 0 & -$a_{2}$ & $a_{2}$ 
& -$a_{2}$ & -$a_{4}$ & -$a_{7}$ & 0 \\
$s_{12}s_{31}u_{12}u_{32}$ 
& $a_{3}$ & -$a_{8}$ & $a_{7}$ & 0 
& $a_{10}$ & $a_{8}$ & -$a_{2}$ & -$a_{8}$ 
& 0 & $a_{1}$ & $a_{8}$ & -$a_{1}$ \\
$s_{12}s_{31}u_{12}u_{31}$ 
& $a_{5}$ & -$a_{8}$ & -$a_{10}$ & -$a_{1}$ 
& 0 & $a_{2}$ & 0 & -$a_{2}$ 
& -$a_{5}$ & $a_{2}$ & $a_{8}$ & $a_{1}$ \\
$s_{12}s_{31}u_{12}u_{23}$ 
& $a_{2}$ & $a_{9}$ & -$a_{2}$ & $a_{1}$  
& 0 & -$a_{2}$ & 0 & $a_{2}$ 
& 0 & -$a_{5}$ & -$a_{9}$ & $a_{14}$ \\
$s_{12}s_{31}u_{12}u_{21}$ 
& $a_{1}$ & -$a_{8}$ & -$a_{10}$ & 0 
& -$a_{10}$ & -$a_{8}$ & $a_{1}$ & $a_{8}$ 
& -$a_{1}$ & 0 & $a_{8}$ & $a_{14}$ \\
$s_{12}s_{31}u_{12}u_{13}$ 
& -$a_{1}$ & $a_{10}$ & $a_{10}$ & 0 
& $a_{2}$ & -$a_{2}$ & -$a_{1}$ & $a_{2}$ 
& 0 & -$a_{5}$ & -$a_{10}$ & -$a_{5}$ \\
$s_{12}s_{31}u_{12}^2$ 
& 0 & $a_{7}$ & $a_{11}$ & -$a_{2}$ 
& $a_{2}$ & 0 & -$a_{2}$ & 0 
& 0 & 0 & -$a_{7}$ & -$a_{4}$ \\
$s_{12}s_{31}^2u_{32}$ 
& $a_{5}$ & $a_{7}$ & -$a_{2}$ & -$a_{2}$ 
& $a_{10}$ & $a_{8}$ & -$a_{2}$ & -$a_{8}$ 
& 0 & -$a_{2}$ & $a_{8}$ & 0 \\
$s_{12}s_{31}^2u_{31}$ 
& $a_{1}$ & -$a_{7}$ & -$a_{11}$ & -$a_{2}$ 
& -$a_{8}$ & $a_{10}$ & $a_{2}$ & -$a_{8}$ 
& -$a_{5}$ & $a_{10}$ & $a_{7}$ & $a_{4}$ \\
$s_{12}s_{31}^2u_{23}$ 
& $a_{1}$ & -$a_{11}$ & -$a_{7}$ & $a_{1}$ 
& $a_{8}$ & -$a_{10}$ & -$a_{1}$ & $a_{10}$ 
& 0 & $a_{5}$ & -$a_{7}$ & $a_{4}$ \\
$s_{12}s_{31}^2u_{21}$ 
& $a_{1}$ & -$a_{9}$ & -$a_{10}$ & 0 
& -$a_{8}$ & $a_{8}$ & 0 & $a_{8}$ 
& -$a_{5}$ & $a_{5}$ & 0 & $a_{6}$ \\
$s_{12}s_{31}^2u_{13}$ 
& 0 & 0 & $a_{9}$ & 0 
& $a_{2}$ & -$a_{2}$ & -$a_{1}$ & $a_{2}$ 
& 0 & 0 & -$a_{9}$ & -$a_{1}$ \\
$s_{12}s_{31}^2u_{12}$ 
& $a_{1}$ & 0 & $a_{9}$ & -$a_{1}$ 
& $a_{1}$ & 0 & -$a_{1}$ & 0 
& 0 & 0 & 0 & -$a_{4}$ \\

\hline\hline
\end{tabular}
\end{center}
\end{table}

\begin{table}
\centerline{Table 1. Continued.}
\begin{center}
\begin{tabular}{lllllllllllll}
\hline\hline
& $A$ & $B$ & $C$ & $D$  & $E$  & $F$ 
& $G$ & $H$ & $I$ & $J$ & $K$ & $L$ \\
\hline
$s_{12}s_{31}^3$ 
& $a_{1}$ & -$a_{7}$ & -$a_{8}$ & -$a_{2}$ 
& $a_{2}$ & 0 & -$a_{2}$ & 0 
& 0 & 0 & $a_{7}$ & 0 \\
$s_{12}s_{23}u_{32}^2$ 
& 0 & $a_{7}$ & $a_{7}$ & $a_{2}$ 
& $a_{8}$ & -$a_{8}$ & -$a_{4}$ & $a_{8}$ 
& -$a_{1}$ & 0 & -$a_{19}$ & 0 \\
$s_{12}s_{23}u_{31}u_{32}$ 
& $a_{2}$ & -$a_{9}$ & $a_{9}$ & 0 
& $a_{10}$ & -$a_{8}$ & -$a_{6}$ & -$a_{8}$ 
& -$a_{2}$ & 0 & 0 & -$a_{10}$ \\
$s_{12}s_{23}u_{31}^2$ 
& $a_{4}$ & -$a_{8}$ & $a_{7}$ & $a_{2}$ 
& $a_{8}$ & $a_{8}$ & 0 & -$a_{8}$ 
& -$a_{2}$ & 0 & $a_{8}$ & -$a_{10}$ \\
$s_{12}s_{23}u_{23}u_{32}$ 
& $a_{1}$ & -$a_{7}$ & -$a_{9}$ & -$a_{1}$ 
& -$a_{8}$ & $a_{8}$ & 0 & -$a_{10}$ 
& $a_{1}$ & $a_{2}$ & $a_{19}$ & $a_{10}$ \\
$s_{12}s_{23}u_{23}u_{31}$ 
& $a_{2}$ & $a_{8}$ & -$a_{2}$ & 0
& -$a_{2}$ & $a_{2}$ & $a_{1}$ & 0 
& 0 & 0 & -$a_{7}$ & $a_{1}$ \\
$s_{12}s_{23}u_{23}^2$ 
& 0 & 0 & $a_{8}$ & 0 
& 0 & 0 & -$a_{2}$ & $a_{2}$ 
& -$a_{2}$ & 0 & 0 & -$a_{8}$ \\
$s_{12}s_{23}u_{21}u_{32}$ 
& $a_{5}$ & -$a_{11}$ & -$a_{7}$ & -$a_{1}$ 
& $a_{2}$ & 0 & -$a_{5}$ & -$a_{2}$ 
& 0 & $a_{2}$ & $a_{7}$ & 0 \\
$s_{12}s_{23}u_{21}u_{31}$ 
& $a_{3}$ & 0 & -$a_{9}$ & 0 
& $a_{8}$ & $a_{10}$ & 0 & -$a_{8}$ 
& -$a_{2}$ & 0 & $a_{20}$ & 0 \\
$s_{12}s_{23}u_{21}u_{23}$ 
& -$a_{1}$ & $a_{9}$ & $a_{7}$ & 0 
& -$a_{2}$ & $a_{2}$ & 0 & $a_{2}$ 
& -$a_{1}$ & -$a_{2}$ & -$a_{22}$ & 0 \\
$s_{12}s_{23}u_{21}^2$ 
& 0 & $a_{9}$ & -$a_{7}$ & -$a_{2}$ 
& 0 & $a_{2}$ & 0 & 0 
& -$a_{2}$ & -$a_{2}$ & $a_{7}$ & 0 \\
$s_{12}s_{23}u_{13}u_{32}$ 
& $a_{2}$ & $a_{8}$ & -$a_{7}$ & -$a_{6}$ 
& 0 & $a_{2}$ & 0 & 0 
& -$a_{1}$ & $a_{15}$ & -$a_{20}$ & $a_{4}$ \\
$s_{12}s_{23}u_{13}u_{31}$ 
& 0 & 0 & 0 & 0 
& -$a_{2}$ & 0 & 0 & 0 
& 0 & $a_{18}$ & -$a_{9}$ & $a_{5}$ \\
$s_{12}s_{23}u_{13}u_{23}$ 
& $a_{3}$ & -$a_{11}$ & -$a_{7}$ & 0 
& $a_{8}$ & -$a_{8}$ & 0 & $a_{10}$ 
& -$a_{2}$ & $a_{4}$ & $a_{21}$ & $a_{2}$ \\
$s_{12}s_{23}u_{13}u_{21}$ 
& $a_{2}$ & -$a_{12}$ & 0 & $a_{1}$ 
& -$a_{2}$ & 0 & 0 & $a_{2}$ 
& 0 & $a_{6}$ & -$a_{7}$ & $a_{1}$ \\
$s_{12}s_{23}u_{13}^2$ 
& $a_{4}$ & $a_{8}$ & -$a_{7}$ & 0 
& $a_{8}$ & -$a_{8}$ & $a_{2}$ & $a_{8}$ 
& -$a_{2}$ & -$a_{4}$ & $a_{7}$ & 0 \\
$s_{12}s_{23}u_{12}u_{32}$ 
& $a_{1}$ & 0 & $a_{7}$ & -$a_{1}$ 
& 0 & $a_{2}$ & -$a_{1}$ & $a_{2}$ 
& -$a_{6}$ & $a_{17}$ & -$a_{21}$ & $a_{15}$ \\
$s_{12}s_{23}u_{12}u_{31}$ 
& $a_{2}$ & -$a_{11}$ & -$a_{7}$ & 0 
& 0 & 0 & -$a_{6}$ & $a_{2}$ 
& -$a_{1}$ & $a_{18}$ & -$a_{8}$ & $a_{3}$ \\
$s_{12}s_{23}u_{12}u_{23}$ 
& $a_{5}$ & -$a_{8}$ & -$a_{2}$ & -$a_{5}$ 
& $a_{2}$ & -$a_{2}$ & -$a_{1}$ & 0 
& 0 & -$a_{1}$ & $a_{7}$ & $a_{2}$ \\
$s_{12}s_{23}u_{12}u_{21}$ 
& $a_{1}$ & -$a_{10}$ & -$a_{8}$ & 0 
& -$a_{8}$ & -$a_{10}$ & -$a_{1}$ & $a_{8}$ 
& $a_{1}$ & $a_{16}$ & $a_{7}$ & $a_{1}$ \\
$s_{12}s_{23}u_{12}u_{13}$ 
& $a_{2}$ & $a_{1}$ & $a_{7}$ & -$a_{6}$ 
& $a_{10}$ & -$a_{8}$ & 0 & -$a_{8}$ 
& -$a_{2}$ & -$a_{17}$ & -$a_{21}$ & -$a_{2}$ \\
$s_{12}s_{23}u_{12}^2$ 
& 0 & $a_{9}$ & $a_{9}$ & -$a_{4}$ 
& $a_{8}$ & $a_{8}$ & $a_{2}$ & -$a_{8}$ 
& -$a_{1}$ & -$a_{10}$ & -$a_{8}$ & -$a_{8}$ \\
$s_{12}s_{23}s_{31}u_{32}$ 
& $a_{5}$ & -$a_{7}$ & -$a_{10}$ & 0 
& $a_{2}$ & 0 & -$a_{5}$ & $a_{2}$ 
& -$a_{6}$ & $a_{16}$ & -$a_{7}$ & $a_{6}$ \\
$s_{12}s_{23}s_{31}u_{31}$ 
& $a_{5}$ & -$a_{12}$ & -$a_{11}$ & 0 
& 0 & $a_{2}$ & -$a_{6}$ & $a_{2}$ 
& -$a_{5}$ & $a_{18}$ & 0 & $a_{17}$ \\
$s_{12}s_{23}s_{31}u_{23}$ 
& $a_{5}$ & -$a_{10}$ & -$a_{7}$ & -$a_{6}$ 
& $a_{2}$ & 0 & -$a_{5}$ & $a_{2}$ 
& 0 & $a_{6}$ & -$a_{7}$ & $a_{16}$ \\
$s_{12}s_{23}s_{31}u_{21}$ 
& $a_{5}$ & -$a_{11}$ & -$a_{12}$ & -$a_{5}$ 
& $a_{2}$ & $a_{2}$ & -$a_{6}$ & 0 
& 0 & $a_{17}$ & 0 & $a_{18}$ \\
$s_{12}s_{23}s_{31}u_{13}$ 
& $a_{5}$ & $a_{7}$ & -$a_{8}$ & -$a_{6}$ 
& 0 & $a_{2}$ & 0 & $a_{2}$ 
& -$a_{5}$ & $a_{14}$ & -$a_{21}$ & $a_{4}$ \\
$s_{12}s_{23}s_{31}u_{12}$ 
& $a_{5}$ & -$a_{8}$ & $a_{7}$ & -$a_{5}$ 
& $a_{2}$ & $a_{2}$ & 0 & 0 
& -$a_{6}$ & $a_{4}$ & -$a_{21}$ & $a_{14}$ \\
$s_{12}s_{23}s_{31}^2$ 
& $a_{6}$ & -$a_{12}$ & -$a_{13}$ & -$a_{2}$ 
& $a_{8}$ & $a_{8}$ & -$a_{2}$ & $a_{8}$ 
& -$a_{5}$ & $a_{15}$ & $a_{22}$ & $a_{6}$ \\
$s_{12}s_{23}^2u_{32}$ 
& $a_{1}$ & -$a_{2}$ & 0 & -$a_{2}$ 
& $a_{10}$ & -$a_{8}$ & -$a_{5}$ & -$a_{8}$ 
& $a_{2}$ & $a_{2}$ & $a_{19}$ & 0 \\
$s_{12}s_{23}^2u_{31}$ 
& $a_{5}$ & -$a_{2}$ & -$a_{7}$ & -$a_{2}$ 
& $a_{8}$ & $a_{10}$ & 0 & -$a_{8}$ 
& -$a_{2}$ & 0 & $a_{2}$ & -$a_{2}$ \\
$s_{12}s_{23}^2u_{23}$ 
& 0 & $a_{8}$ & -$a_{8}$ & 0 
& -$a_{2}$ & $a_{2}$ & 0 & $a_{2}$ 
& -$a_{1}$ & 0 & -$a_{7}$ & $a_{2}$ \\
$s_{12}s_{23}^2u_{21}$ 
& $a_{1}$ & $a_{7}$ & -$a_{9}$ & -$a_{1}$ 
& 0 & $a_{1}$ & 0 & 0 
& -$a_{1}$ & -$a_{2}$ & $a_{9}$ & 0 \\
$s_{12}s_{23}^2u_{13}$ 
& $a_{1}$ & -$a_{8}$ & -$a_{2}$ & $a_{1}$ 
& -$a_{10}$ & $a_{8}$ & 0 & $a_{10}$ 
& -$a_{1}$ & $a_{6}$ & -$a_{8}$ & $a_{5}$ \\
$s_{12}s_{23}^2u_{12}$ 
& $a_{1}$ & -$a_{11}$ & 0 & 0 
& $a_{8}$ & -$a_{8}$ & -$a_{5}$ & $a_{8}$ 
& 0 & $a_{6}$ & -$a_{21}$ & $a_{4}$ \\
$s_{12}s_{23}^2s_{31}$ 
& $a_{6}$ & -$a_{13}$ & -$a_{13}$ & -$a_{2}$ 
& $a_{8}$ & $a_{8}$ & -$a_{5}$ & $a_{8}$ 
& -$a_{2}$ & $a_{6}$ & $a_{22}$ & $a_{6}$ \\
$s_{12}s_{23}^3$ 
& $a_{1}$ & -$a_{8}$ & -$a_{8}$ & -$a_{2}$ 
& 0 & $a_{2}$ & 0 & 0 
& -$a_{2}$ & 0 & $a_{8}$ & 0 \\
$s_{12}^2u_{32}^2$ 
& 0 & $a_{8}$ & -$a_{8}$ & 0 
& $a_{8}$ & -$a_{8}$ & 0 & -$a_{8}$ 
& $a_{1}$ & $a_{2}$ & -$a_{7}$ & $a_{1}$ \\
$s_{12}^2u_{31}u_{32}$ 
& -$a_{1}$ & $a_{7}$ & 0 & -$a_{5}$ 
& $a_{8}$ & $a_{8}$ & $a_{2}$ & -$a_{10}$ 
& $a_{2}$ & $a_{3}$ & -$a_{9}$ & $a_{5}$ \\
$s_{12}^2u_{31}^2$ 
& 0 & -$a_{8}$ & $a_{7}$ & 0 
& -$a_{8}$ & $a_{8}$ & $a_{1}$ & -$a_{8}$ 
& 0 & $a_{5}$ & -$a_{7}$ & $a_{2}$ \\
$s_{12}^2u_{23}u_{32}$ 
& $a_{1}$ & -$a_{8}$ & $a_{8}$ & 0 
& -$a_{8}$ & $a_{8}$ & 0 & $a_{8}$ 
& 0 & 0 & $a_{7}$ & -$a_{2}$ \\
$s_{12}^2u_{23}u_{31}$ 
& $a_{5}$ & -$a_{8}$ & -$a_{7}$ & 0 
& 0 & 0 & 0 & $a_{2}$ 
& 0 & 0 & $a_{7}$ & 0 \\

\hline\hline
\end{tabular}
\end{center}
\end{table}

\begin{table}
\centerline{Table 1. Continued.}
\begin{center}
\begin{tabular}{lllllllllllll}
\hline\hline
& $A$ & $B$ & $C$ & $D$  & $E$  & $F$ 
& $G$ & $H$ & $I$ & $J$ & $K$ & $L$ \\
\hline
$s_{12}^2u_{21}u_{32}$ 
& 0 & -$a_{9}$ & $a_{7}$ & -$a_{5}$ 
& 0 & $a_{2}$ & 0 & 0 
& -$a_{1}$ & $a_{4}$ & -$a_{7}$ & $a_{10}$ \\
$s_{12}^2u_{21}u_{31}$ 
& $a_{1}$ & -$a_{2}$ & 0 & 0 
& -$a_{8}$ & $a_{8}$ & $a_{2}$ & $a_{8}$ 
& -$a_{2}$ & $a_{4}$ & 0 & $a_{8}$ \\
$s_{12}^2u_{13}u_{32}$ 
& $a_{5}$ & -$a_{7}$ & 0 & 0 
& 0 & 0 & 0 & $a_{2}$ 
& 0 & -$a_{2}$ & 0 & 0 \\
$s_{12}^2u_{13}u_{31}$ 
& $a_{1}$ & $a_{8}$ & -$a_{7}$ & 0 
& $a_{8}$ & -$a_{8}$ & 0 & $a_{8}$ 
& 0 & -$a_{1}$ & $a_{7}$ & 0 \\
$s_{12}^2u_{12}u_{32}$ 
& $a_{1}$ & $a_{7}$ & -$a_{8}$ & 0 
& $a_{8}$ & -$a_{8}$ & -$a_{2}$ & $a_{8}$ 
& $a_{2}$ & -$a_{1}$ & -$a_{7}$ & 0 \\
$s_{12}^2u_{12}u_{31}$ 
& 0 & $a_{8}$ & -$a_{7}$ & -$a_{5}$ 
& $a_{2}$ & 0 & -$a_{1}$ & 0 
& 0 & -$a_{1}$ & -$a_{8}$ & $a_{2}$ \\
$s_{12}^2s_{31}u_{32}$ 
& $a_{1}$ & -$a_{7}$ & -$a_{11}$ & 0 
& $a_{10}$ & -$a_{10}$ & -$a_{1}$ & $a_{8}$ 
& $a_{1}$ & $a_{4}$ & -$a_{7}$ & $a_{5}$ \\
$s_{12}^2s_{31}u_{31}$ 
& $a_{1}$ & -$a_{10}$ & -$a_{9}$ & -$a_{5}$ 
& $a_{8}$ & $a_{8}$ & 0 & -$a_{8}$ 
& 0 & $a_{6}$ & 0 & $a_{5}$ \\
$s_{12}^2s_{31}u_{23}$ 
& $a_{5}$ & -$a_{2}$ & $a_{7}$ & 0 
& -$a_{8}$ & $a_{8}$ & -$a_{2}$ & $a_{10}$ 
& -$a_{2}$ & 0 & $a_{8}$ & -$a_{2}$ \\
$s_{12}^2s_{31}u_{21}$ 
& $a_{1}$ & -$a_{11}$ & -$a_{7}$ & -$a_{5}$ 
& -$a_{8}$ & $a_{10}$ & $a_{2}$ & -$a_{8}$ 
& -$a_{2}$ & $a_{4}$ & $a_{7}$ & $a_{10}$ \\
$s_{12}^2s_{31}u_{13}$ 
& $a_{1}$ & $a_{9}$ & 0 & 0 
& 0 & 0 & -$a_{1}$ & $a_{1}$ 
& -$a_{1}$ & -$a_{4}$ & 0 & 0 \\
$s_{12}^2s_{31}u_{12}$ 
& 0 & $a_{9}$ & 0 & 0 
& $a_{2}$ & -$a_{2}$ & -$a_{1}$ & $a_{2}$ 
& 0 & -$a_{1}$ & -$a_{9}$ & 0 \\
$s_{12}^2s_{31}^2$ 
& $a_{1}$ & -$a_{9}$ & -$a_{9}$ & 0 
& $a_{2}$ & -$a_{2}$ & -$a_{1}$ & $a_{2}$ 
& 0 & $a_{1}$ & 0 & $a_{1}$ \\
$s_{12}^2s_{23}u_{32}$ 
& $a_{1}$ & -$a_{8}$ & -$a_{2}$ & -$a_{5}$ 
& $a_{8}$ & $a_{8}$ & 0 & -$a_{8}$ 
& 0 & $a_{3}$ & -$a_{22}$ & $a_{15}$ \\
$s_{12}^2s_{23}u_{31}$ 
& $a_{1}$ & -$a_{10}$ & -$a_{8}$ & 0 
& -$a_{10}$ & $a_{10}$ & $a_{1}$ & $a_{8}$ 
& -$a_{1}$ & $a_{18}$ & -$a_{7}$ & $a_{4}$ \\
$s_{12}^2s_{23}u_{23}$ 
& $a_{1}$ & -$a_{8}$ & $a_{7}$ & 0 
& 0 & 0 & -$a_{1}$ & $a_{1}$ 
& -$a_{1}$ & 0 & $a_{7}$ & -$a_{8}$ \\
$s_{12}^2s_{23}u_{21}$ 
& 0 & -$a_{9}$ & 0 & 0 
& -$a_{2}$ & $a_{2}$ & 0 & $a_{2}$ 
& -$a_{1}$ & $a_{4}$ & -$a_{19}$ & $a_{2}$ \\
$s_{12}^2s_{23}u_{13}$ 
& $a_{5}$ & 0 & -$a_{8}$ & 0 
& $a_{8}$ & -$a_{8}$ & -$a_{2}$ & $a_{10}$ 
& -$a_{2}$ & -$a_{1}$ & $a_{8}$ & 0 \\
$s_{12}^2s_{23}u_{12}$ 
& $a_{1}$ & $a_{7}$ & -$a_{7}$ & -$a_{5}$ 
& $a_{10}$ & -$a_{8}$ & -$a_{2}$ & -$a_{8}$ 
& $a_{2}$ & -$a_{1}$ & -$a_{22}$ & $a_{8}$ \\
$s_{12}^2s_{23}s_{31}$ 
& $a_{6}$ & -$a_{13}$ & -$a_{12}$ & -$a_{5}$ 
& $a_{8}$ & $a_{8}$ & -$a_{2}$ & $a_{8}$ 
& -$a_{2}$ & $a_{6}$ & $a_{22}$ & $a_{15}$ \\
$s_{12}^2s_{23}^2$ 
& $a_{1}$ & -$a_{2}$ & -$a_{9}$ & 0 
& -$a_{2}$ & $a_{2}$ & 0 & $a_{2}$ 
& -$a_{1}$ & $a_{5}$ & 0 & $a_{1}$ \\
$s_{12}^3u_{32}$ 
& $a_{1}$ & -$a_{8}$ & -$a_{7}$ & 0 
& 0 & 0 & -$a_{2}$ & $a_{2}$ 
& -$a_{2}$ & 0 & $a_{7}$ & 0 \\
$s_{12}^3u_{31}$ 
& $a_{1}$ & -$a_{8}$ & -$a_{7}$ & 0 
& 0 & 0 & -$a_{2}$ & $a_{2}$ 
& -$a_{2}$ & 0 & $a_{7}$ & 0 \\
$s_{12}^3s_{31}$ 
& $a_{1}$ & -$a_{8}$ & -$a_{7}$ & 0 
& 0 & 0 & -$a_{2}$ & $a_{2}$ 
& -$a_{2}$ & 0 & $a_{7}$ & 0 \\
$s_{12}^3s_{23}$ 
& $a_{1}$ & -$a_{8}$ & -$a_{7}$ & 0 
& 0 & 0 & -$a_{2}$ & $a_{2}$ 
& -$a_{2}$ & 0 & $a_{7}$ & 0 \\

\hline\hline
\end{tabular}
\end{center}
\end{table}

\end{document}